\documentclass[11pt]{article}
\usepackage{graphicx}
\usepackage{natbib}
\usepackage[table]{xcolor}
\definecolor{lightgray}{gray}{0.9}
\setcitestyle{authoryear,open={(},close={)}}

\begin{document}

\title{ \bf Hierarchical networks of scientific journals}
\author{Gergely Palla$^{1,2}$\footnote{Corresponding author, e-mail: pallag@hal.elte.hu}, Gergely Tib{\'e}ly$^3$, Enys Mones$^2$, P{\'e}ter Pollner$^{1,2}$\\ and Tam{\'as} Vicsek$^{1,3}$\\ 
\footnotesize{$^1$MTA-ELTE Statistical and Biological Physics Research Group,}\\
\footnotesize{Hungarian Academy of Sciences, }\\
\footnotesize{P{\'a}zm{\'a}ny P. stny 1/A, H-1117 Budapest, Hungary,}\\ 
\footnotesize{$^2$Regional Knowledge Centre, E{\"o}tv{\"os} University, H-8000 Sz{\'e}kesfeh{\'e}rv{\'a}r, Hungary,}\\
\footnotesize{$^3$Dept. of Biological Physics, E{\"o}tv{\"os} University,}\\
\footnotesize{ P{\'a}zm{\'a}ny P. stny 1/A, H-1117 Budapest, Hungary}}
\maketitle

\begin{abstract}
Scientific journals are the repositories of mankind’s gradually accumulating knowledge of the surrounding world. Just as knowledge is organised into classes ranging from major disciplines, subjects and fields to increasingly specific topics, journals can also be categorised into groups using various metrics. In addition, they can be ranked according to their overall influence. However, according to recent studies, the impact, prestige and novelty of journals cannot be characterised by a single parameter such as, e.g., the impact factor. In order to deepen our insight into the impact of a journal, the knowledge gap our work is intended to fill is the evaluation of journal relevance using complex multi-dimensional measures. Thus, for the first time, our objective is to organise journals into multiple hierarchies based on citation data. The two approaches we use are designed to address this problem from different perspectives. We use a measure related to the notion of m-reaching centrality and find a network that shows a journal’s level of influence in terms of the direction and efficiency with which information spreads through the network. We find we can also obtain an alternative network using a suitably modified nested hierarchy extraction method applied to the same data. In this case, in a self-organized way, the journals become branches according to the major scientific fields, where the local structure of the branches is reflecting the hierarchy within the given field, with usually the most prominent journal (according to other measures) in the field chosen by the algorithm as the local root, and more specialised journals positioned deeper in the branch. This can make the navigation within different scientific fields and sub-fields very simple, being equivalent to navigating in the different branches of the nested hierarchy and, for example, should be very helpful when choosing the most appropriate journal for a given manuscript. According to our results, the two alternative hierarchies show a somewhat different but also consistent picture of the intricate relations between scientific journals, and, as such, they also display a new angle of view on how our scientific knowledge is organized into networks.
\end{abstract}

\section*{Introduction}

Providing an objective ranking of scientific journals and mapping them into different knowledge domains are very complex problems of significant importance, which can be achieved using a number of different approaches. Probably the most widely known quality measure is the impact factor \citep{Garfield_impact,Garfield_impact_2}, corresponding to the total number of citations a journal receives in a 2-year period, divided by the number of published papers over the same period. Although it is a rather intuitive quantity, the impact factor is also suffering from serious limitations \citep{Opthof_IF_nonsense,Seglen_IF_nonsense,Harter_IF_nonsense,Bordons_IF_nonsense}. This consequently led to the introduction of alternative measures such as the H-index for journals \citep{Braun_H_index}, the g-index \citep{Egghe_g_index}, the Eigenfactor \citep{Bergstrom_impact}, the PageRank and the Y-factor \citep{Bollen_impact}, the Scimago Journal Rank \citep{scimago}, and the use of various centralities such as the degree-, closeness- or betweeness centrality in the citation network between the journals \citep{Bollen_centralities,Leydesdorff_betweenness}. Comparing the advantages and disadvantages of the different impact measures and examining their correlation has attracted considerable interest in the literature \citep{Bollen_Plos,Franceschet_biblio,Franceschet_reasons,Glanzel_impact,Filippo_impact}. However, according to the results of the principal component analysis of 39 quality measures carried out by \cite{Bollen_Plos}, scientific impact is a multi-dimensional construct that cannot be adequately measured by any single indicator. Thus, the development of higher dimensional quality indicators for scientific journals provides an important objective for current research.

In this study we consider different possibilities for defining a hierarchy between scientific journals based on their citation network. The advantage of using the network approach for representing the intricate relations between journals is that networks can show  substantially different aspects compared to any parametric method representing the journals with points in single- or even in multi-dimensional space. When organised into a hierarchy, the most important and prestigious journals are expected to appear at the top, while lesser known journals are expected to be ranked lower. However, a hierarchy offers a more complex view of the ranking between journals compared to a one dimensional impact measure. For example, if the branches of the hierarchy are organised according to the different scientific fields, then the journals in a given field can be compared simply by zooming into the corresponding branch in the hierarchy. Possible scenarios for hierarchical relations between scientific journals have already been suggested by Iyengar and Balijepally recently \citep{dominance_hierarchy}. However, the main objective in this earlier study \citep{dominance_hierarchy} was to examine the validity of a linear ordering between the journals based on a dominance ranking procedure. Here we construct and visualise multiple hierarchies between the journals, offering a far more complex view of the ranking between journals compared to a one dimensional impact measure.

Hierarchical organisation in general is a widespread phenomenon in nature and society. This is supported by several studies, focusing on the transcriptional regulatory network of Escherichia coli \citep{Zeng_Ecoli}, the dominant-subordinate hierarchy among crayfish \citep{Huber_crayfish}, the leader-follower network of pigeon flocks \citep{Tamas_pigeons}, the rhesus macaque kingdoms \citep{McCowan_macaque}, neural networks \citep{Kaiser_neural}, technological networks \citep{Pumain_book}, social interactions \citep{Guimera_hier_soc,our_pref_coms,Sole_hier_soc}, urban planning \citep{Krugman_urban,Batty_urban}, ecological systems \citep{Hirata_eco,Wickens_eco}, and evolution \citep{Eldrege_book,McShea_organism}. However, hierarchy is a polysemous word, and in general, we can distinguish between three different types of hierarchies when describing a complex system: the order, the nested and the flow hierarchy. In the case of order hierarchy, we basically define a ranking, or more precisely a partial ordering, of the set of elements under investigation \citep{Lane_order_hierarchy}. Nested hierarchy, (also called as inclusion hierarchy or containment hierarchy), represents the idea of recursively aggregating the items into larger and larger groups, resulting in a structure where higher level groups consist of smaller and more specific components \citep{Wimberley_book}. Finally, a flow hierarchy can be depicted as a directed graph, where the nodes are layered in different levels so that the nodes that are influenced by a given node (are connected to it through a directed link) are at lower levels. 

Hierarchical organisation is an important concept also in network theory \citep{Laci_hier_scale,Sneppen_hier_measures,Newman_hier,Pumain_book,Sole_chaos_hier,Enys_hierarchy,Sole_hier_PNAS}. The network approach has become an ubiquitous tool for analysing complex systems - from the interactions within cells, transportation systems, the Internet and other technological networks, through to economic networks, collaboration networks and society \citep{Laci_revmod,Dorog_book}. Grasping the signs of hierarchy in networks is a non-trivial task with a number of possible different approaches, including the statistical inference of an underlying hierarchy based on the observed network structure \citep{Newman_hier}, and the introduction of various hierarchy measures \citep{Sneppen_hier_measures,Enys_hierarchy,Sole_hier_PNAS}. What makes the analysis of hierarchy even more complex is that it may also be context dependent. According to a recent study on homing pigeons, the hierarchical pattern of in-flight leadership does not build upon the stable, hierarchical social dominance structure (pecking order) evident among the same birds \citep{Pigeon_context}. 

In this study we show that in a somewhat similar fashion, scientific journals can also be organised into multiple hierarchies with different types. Our studies rely on the citation network between scientific papers obtained from the Web of Science \citep{WOS}. 
On the one hand, the flow hierarchy analysis of this network based on the $m$-reaching centrality \citep{Enys_hierarchy,Borgatti_reach} reveals the structure relevant from the point of view of knowledge spreading and influence. On the other hand, the alternative hierarchy obtained from the same network with the help of an automated tag hierarchy extraction method \citep{Tibely_plosone} highlights a nested structure with the most interdisciplinary journals at the top and the very specialised journals at the bottom of the hierarchy. 

\section*{Scientific publication data}
The data set on which our studies rely consists of all the available publications in the Web of Science \citep{WOS} between 1975 and 2011. The downloading scripts we used are available in \citep{wos_scripts}. In order to take into account a list of papers as wide as possible, we did not apply any specific filtering. Thus, conference proceedings and technical papers also appear in the used data set. However, since the network we study builds upon citation between papers (or journals), the conference proceedings, technical papers (or even journals) with no incoming citation fall out of the flow hierarchy analysis automatically. (Nevertheless, in the event that they have outgoing citations, this is included in the evaluation of the $m$-reaching centrality of other journals). Furthermore, even when cited, a conference proceedings does not have a real chance of getting high in any of the hierarchies considered here, due to their very limited number of publications compared to journals. Although highly cited individual conference proceedings publications may appear, they cannot boost the overall citation of the proceedings to the level of journals, (e.g., whenever a scientific breakthrough is published in a conference proceedings first, it is usually also published in a more prestigious journal soon afterwards, which eventually drives the citations to the journal instead of the proceedings). For these reasons, the conference proceedings are ranked at the bottom of the hierarchies we obtained.

We used the 11 character-long abbreviated journal issue field in the core data for identifying the journal of a given publication. The advantage of using this field is that it contains only an abbreviated journal name without any volume numbers, issue numbers, years, and so on, (in contrast, the full journal name in some cases may contain the volume number or the publication year as well, which of course, are varying over time). The total number of publications for which the mentioned data field was non-empty reached 35,372,038, and the number of different journals identified based on this data field was 13,202. As mentioned previously, in case of conference proceedings, the appearing 11 character long abbreviated journals issue field was treated the same as in case of journal publications, without any filtering.

\section*{Flow hierarchy based on the m-reaching centrality}
\label{Sect:GRC}
A recently introduced approach for quantifying the position of a node in a flow hierarchy is based on the $m$-reaching centrality \citep{Enys_hierarchy}. The basic intuition behind this idea is that reaching the rest of the network should be relatively easy for the nodes high in the hierarchy, and more difficult from the nodes at the bottom of the hierarchy. Thus, the position of the node $i$ in the hierarchy is determined by its $m$-reaching centrality \citep{Borgatti_reach}, $C_{m}(i)$, corresponding to the fraction of nodes that can be reached from $i$, following directed paths of at most $m$ steps, (where $m$ is a system dependent parameter). Naturally, a higher $C_{m}(i)$ value corresponds to a higher position in the hierarchy, and the node with the maximal $C_{m}(i)$ is chosen as the root. However, this approach does not specify the ancestors or descendants of a given node in the hierarchy, instead it provides only a ranking between the nodes of the underlying network according to  $C_{m}(i)$. Nevertheless, hierarchical levels can be defined in a simple way: after sorting nodes in an ascending order, we can sample and aggregate nodes into levels so that in each level, the standard deviation of $C_m$ is lower than a pre-defined fraction of the standard deviation in the whole network. This method of constructing a flow hierarchy based on the $m$-reach (and the standard variation of the $m$-reach) has already been shown to provide meaningful structures for a couple of real systems including electric circuits, transcriptional regulatory networks, e-mail networks and food webs \citep{Enys_hierarchy}.

When applying this approach to the study of the hierarchy between scientific journals, we have to take into account that journals are not directly connected to each other; instead they are linked via a citation network between the individual publications. In principle we may assume different `journal strategies' for obtaining a large reach in this system: for example, a journal might publish a very high number of papers of poor quality with only a few citations each. Nevertheless, taken together they can still provide a large number of aggregated citations. Another option is to publish a lower number of high-quality papers, obtaining a lot of citations individually. To avoid having a built-in preference for one type of journal over the other, we define a reaching centrality that is not sensitive to such details, and which only depends on the number of papers that can be reached in $m$-steps from publications appearing in a given journal.

\begin{figure}[h!]
\centerline{\includegraphics[width=0.85\textwidth]{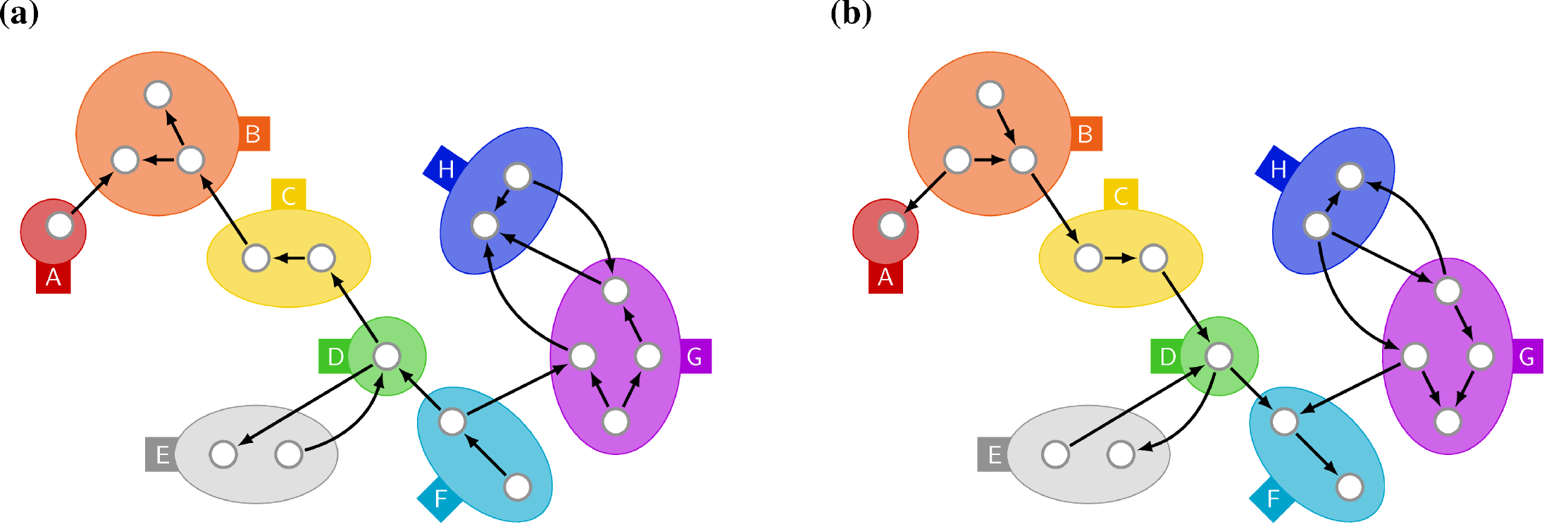}}
\caption{
A schematic illustration of the calculation of journals' $m$-reach. a) The papers are represented by grey nodes, connected by directed citation links, while the journals are corresponding to the coloured sets. b) When calculating the $m$-reach, the links have to be reverted. For example, in the case of a 3-reach, articles in journal B (orange) can reach 4 papers in total (the ones in journals A, C and D) excluding journal B itself; while journal H (dark blue) has a reach of 6,  (given by articles in journals F and G). Note that if we would switch to network between journals, B would have a reach of 5, (as it can reach journals A, C, D, E and F); while journal H would have a reach of 2, (since it could reach only journals F and G).
}
\label{fig:reach_illustr}
\end{figure}

First, we note that when calculating the reach of the publications, the citation links have to be followed backwards: that is, if paper $i$ is citing $j$, then the information presented in $j$ has reached $i$. Thus, the reaching centralities are evaluated in a network where the links are pointing from a reference article to all papers citing it. The $m$-reach of a journal $\mathcal{J}$, denoted by $C_m(\mathcal{J})$, is naturally given by the number of papers that can be reached in at most $m$ steps from any article appearing in the given journal. Thus, the mathematical definition of $C_m(\mathcal{J})$ is based on the set of $m$-reachable nodes, given by
\begin{equation}
  \mathcal{C}_m(\mathcal{J})=\{ \,i \,\, | \,\, d_\mathrm{out}(j,i) \le m, j \in \mathcal{J} \land i \notin \mathcal{J} \, \},
\label{eq:reach_def}
\end{equation}
where $d_\mathrm{out}(j,i)$ denotes the out-distance from paper $j$ to $i$, (i.e., the distance of the papers when only consecutive out-links are considered). The set $\mathcal{C}_m(\mathcal{J})$ is equivalent to the set of papers outside $\mathcal{J}$ that can be reached in at most $m$ steps, provided that the starting publication is in $\mathcal{J}$. The $m$-reaching centrality of $\mathcal{J}$ is simply the size of the $m$-reachable set,  $C_m(\mathcal{J})=|\mathcal{C}_m(\mathcal{J})|$,  (i.e., the number of papers in $\mathcal{C}_m(\mathcal{J})$). Fig.\ref{fig:reach_illustr}.\ shows an illustration of the calculation of the $m$-reach of the journals detailed above. We note that a closely related impact measure for judging the influence of research papers based on deeper layers of other papers in the citation network is given by the wake-citation-score \citep{wake_citation}. A comparison study between the $m$-reach and the wake-citation-score is given in the Supplementary Information S1.

In order to determine the optimal value of $m$, we calculated the $C_m(\mathcal{J})$ for all journals in our data set for a wide range of $m$ values. According to the results detailed in the Supplementary Information S2., around $m=4$ the $C_m(\mathcal{J})$ starts to saturate for the top journals. In order to provide a fair and robust ordering between the journals, here we set $m$ to $m=3$, corresponding to an optimal setting: on the one hand we are still allowing multiple steps in the paths contributing to the reach. On the other hand, we also avoid the saturation effect caused by the exponential increase in the reach as a function of the maximal path length and the finite system size. (More details on the tuning of $m$ are given in the Supplementary Information S2, and the results obtained for other $m$ values are shown in the Supplementary Information S3).

Before considering the results, we note that an alternative approach for studying the citation between journals is to aggregate all papers in a given journal into a single node, representing the journal itself, in similar fashion to the works by \cite{Leydesdorff_aggregate,Leydesdorff_aggregate_2}. In this case the link weight from journal $\mathcal{J}$ to journal $\mathcal{I}$ is given by total number of citations from papers appearing in $\mathcal{J}$ to papers in $\mathcal{I}$. In the Supplementary Information S4.\ we analyse the flow hierarchy obtained by evaluating the $m$-reaching centrality in this aggregated network between the journals. However, recent works have pointed out that aggregations of this nature can lead to serious misjudgement of the importance of nodes \citep{Rosvall_memory,Ingo_aggregate}. For instance, an interesting memory effect of the citation network between individual papers is that a paper citing mostly biological papers that appear in an interdisciplinary journal, is still much more likely to be cited back by other biological papers compared to other disciplines \citep{Rosvall_memory}. Such phenomena can have a significant influence on the $m$-reaching centrality. However, by switching to the aggregated network between journals we wipe out these effects and introduce a distortion in the $m$-reach. Thus, here we stick to the most detailed representation of the system, given by the citation network between individual papers, and leave the analysis of the aggregated network between journals to the Supplementary Information S4. (An illustration of the difference between the $m$-reach calculated on the level of papers and on the aggregated level of journals is given in Fig.\ref{fig:reach_illustr}.)

\begin{figure}[hbt]
\begin{center}
\centerline{\includegraphics[width=\textwidth]{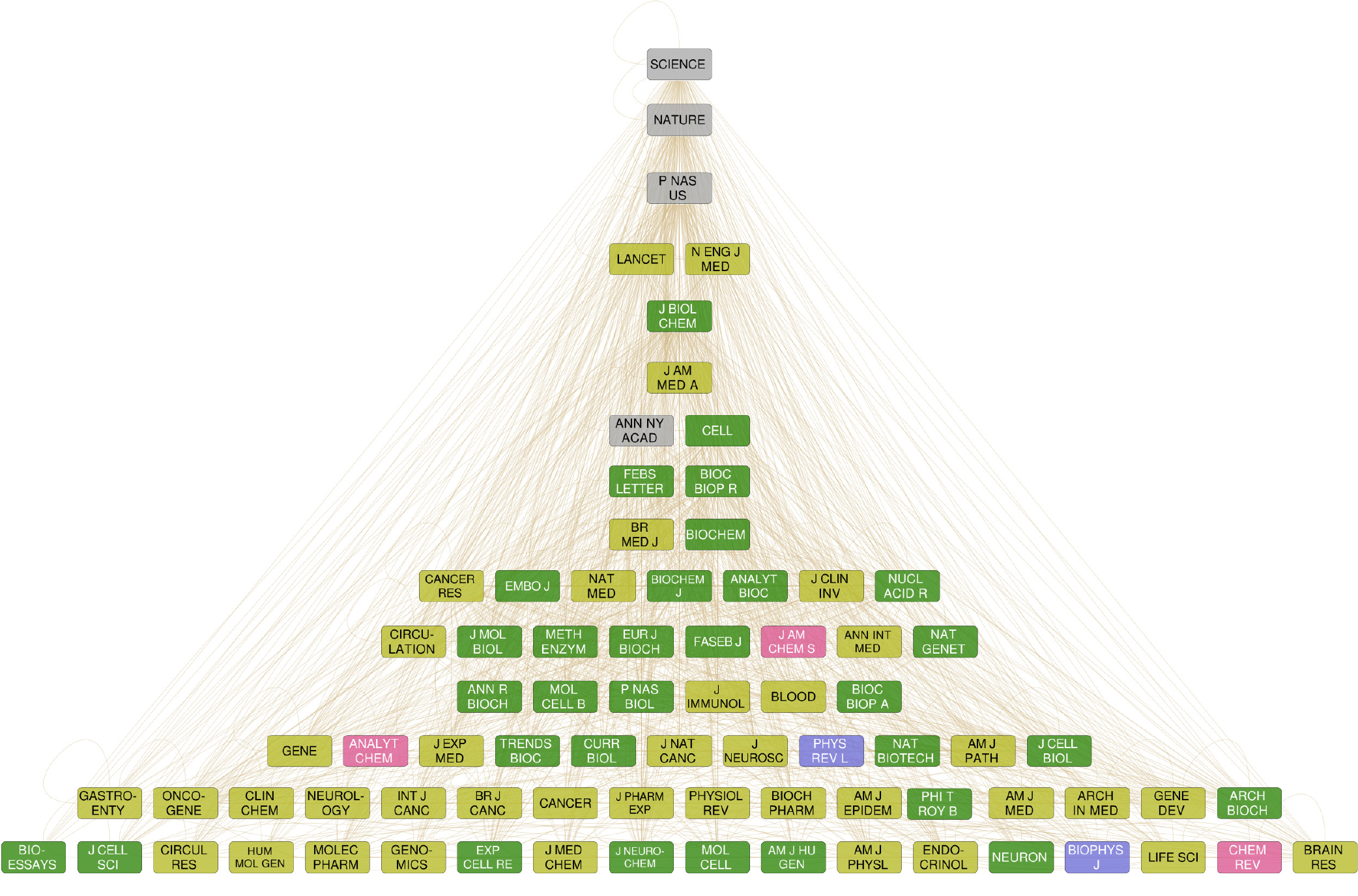}}
\caption{
Top journals in the flow hierarchy according to the $m$-reaching centrality $C_m$ at $m=3$, based on the scientific publication data from the Web of Science. The standard deviation of $C_m$ within the individual hierarchy levels is at most $0.13$$\cdot \sigma(C_m)$, where $\sigma(C_m)$ denotes the standard deviation of $C_m$ over all journals. The nodes are coloured according to the scientific field of the given journal.}
\label{fig:GRC_hierarchy}
\end{center}
\end{figure}

The results for the top journals according to the $m$-reaching centrality at $m=3$ based on the publication data available from the Web of Science between 1975 and 2011 are given in Fig.\ref{fig:GRC_hierarchy}. The hierarchy levels were defined by allowing a maximal standard deviation of $0.13 \cdot \sigma(C_m)$ for $C_m$ within a given level, where $\sigma(C_m)$ denotes the standard deviation of $C_m$ over all journals. (The effect of changes in the within-level standard deviation of $C_m$ on the shape of the hierarchy is discussed in the Supplementary Information S5). According to our analysis, Science is the most influential journal based on the flow hierarchy, followed by Nature, with PNAS coming third, while Lancet and the New England Journal of Medicine forme the fourth level. In general, the top of the hierarchy is strongly dominated by medical, biological and biochemical journals. For instance, the top physics journal, the Physical Review Letters appears only on the 13$^{\rm th}$ level, and the top chemistry journal, the Journal of the American Chemical Society is positioned at the 11$^{\rm th}$ level.

For comparison, in Fig.S7.\ in the Supplementary Information S4.\ we show the top of the flow hierarchy obtained from the citation network aggregated to the level of journals. Although Science, Nature and PNAS preserve their position as the top three journals, relevant changes can be observed in the hierarchy levels just below, as physical and chemical journals take over the biological and medical journals. For instance, Physical Review Letters is raised from level 13 to level three, while Lancet is pushed back from level four to level 17. This reorganisation is likely to be caused by the `memory' of the citation network described in the work by \cite{Rosvall_memory}, - the fact that a paper citing mostly biological articles is more likely to be cited by other biological papers, even if it appears in an interdisciplinary journal. Since biology and medicine have the highest publication rate among different scientific fields, the aggregation to the level of journals has the most severe effect on the reach of entities obtaining citations mostly from these fields. Thus, the notable difference between the flow hierarchy obtained from the citation network of individual papers and from the aggregated network between journals is yet another indication of the distortion in centralities caused by link aggregation, pointed out in related, but somewhat different contexts by \cite{Rosvall_memory} and by \cite{Ingo_aggregate}.

\section*{Extracting a nested hierarchy}
\label{Sect:nested}

Categorising items into a nested hierarchy is a general idea that has been around for a long time in, for instance, library classification systems, biological classification and also in the content classification of scientific publications. A very closely related problem is that given by the automatised categorisation of free tags appearing in various on-line content \citep{Garcia-Molina,Lerman_constr_2,Schmitz_constr,Van_Damme_constr,Tibely_plosone,Nyelveszek}. In the recent years the voluntary tagging of photos, films, books, and so on, with free words has become popular on the internet in blogs, various file sharing platforms  on-line stores and news portals. In some cases these phenomena are referred to as collaborative tagging \citep{Lambiotte_ct,Cattuto_PNAS,Schoder_tags,Cattuto_PNAS2}, and the resultant large collections of tags are referred to as folksonomies, highlighting their collaborative origin and the ``flat'' organisation of the tags in these systems \citep{Lambiotte_ct,Cattuto_PNAS,Cattuto_PNAS2,Mika_folk_and_ont,Spyns_folk_and_ont,Voss_cond_mat,our_ontology}. The natural mathematical representation of tagging systems is given by hypergraphs \citep{Newman_PRE,Caldarelli_PRE}.

 Revealing the hidden hierarchy between tags in a folksonomy or a tagging system in general can significantly help broadening or narrowing the scope of search in the system, give recommendation about yet unvisited objects to the user, or help the categorisation of newly appearing objects \citep{Zhou_recommend_overview,Kazienko_chapter}. Here we apply a generalised version of a recent tag hierarchy extraction method \citep{Tibely_plosone} for constructing a nested hierarchy between scientific journals. In its original form, the input of the tag hierarchy extraction algorithm is given by the weighted co-occurrence network between the tags, where the weights correspond to number of shared objects. Based on the $z$-score of the connected pairs and the centrality of the tags in the co-occurrence network, the hierarchy is built bottom up, as the algorithm eventually assigns one or a few direct ancestors to each tag (except for the root of the hierarchy). The details of the algorithm are described in the Nested hierarchy exctraction algorithm subsection.

In order to study the nested hierarchy between scientific journals, we simply replace the weighted co-occurrence network between tags by the weighted citation network between journals at the input of the algorithm. Although a tag co-occurrence network and a journal citation network are different, the two most important properties needed for the nested hierarchy analysis are the same in both: general tags and multidisciplinary journals have a significantly larger number of neighbours compared to more specific tags and specialised journals. Furthermore, closely related tags co-appear more often compared to unrelated tags, as journals focusing on the same field cite each other more often compared to journals dealing with independent disciplines. Based on this, the hierarchy obtained from the journal citation network in this approach is expected to be organised according to the scope of the journals, with the most general multidisciplinary journals at the top and the very specialised journals at the bottom. 

We note that since in this case we have to determine which journals are the most closely related to each other and which are unrelated, rather than evaluating the overall influence of the journals, we use simply the number of direct citations from one journal to the other as the weight for the connections. This is equivalent of taking the $m$-reach calculated on the publication level at $m=1$, sorting according to the source of the citations and then summing up the results for the papers appearing in one given journal. Thus, when constructing the flow hierarchy, we start from the publication level citation network and evaluate the $m$-reach at $m=3$, whereas in case of the nested hierarchy we calculate the publication level $m$-reach at $m=1$, which technically becomes equivalent to the journal level citation numbers when summed over papers appearing in one given journal.

\subsection*{Nested hierarchy extraction algorithm}

Our algorithm corresponds to a generalised version of ``Algorithm B'' presented in \citep{Tibely_plosone}. The main differences are that here we force the algorithm to produce a directed acyclic graph consisting of a single connected component, and we allow the presence of multiple direct ancestors. In contrast, in its original form ``Algorithm B'' can provide disconnected components, and each component in the output is corresponding to a directed tree. A further technical improvement we introduce is given by the calculation of the node centralities. Thus, the outline of the method used here is the following: first we carry out ``Algorithm B'' given in \citep{Tibely_plosone} with modified centrality evaluation, obtaining a directed tree between the journals. This is followed by a second iteration where we `enrich' the hierarchy by occasionally assigning further direct ancestors to the nodes.

Since `Algorithm B'' is presented in full detail in \citep{Tibely_plosone}, here we provide only a brief overview. The input of the algorithm is a weighted directed network between the journals based on the $z$-score for the citation links. After throwing away unimportant connections by using a weight threshold, the node centralities are evaluated in the remaining network. Here we used a centrality based on random walks on the citation network between journals with occasional teleportation steps, in a similar fashion to PageRank. We adopted the method proposed by \cite{centrality_arxiv}, calculating the dominant right eigenvector of the matrix $M_{ij}=(1-\alpha)w_{ij}+\alpha s_i^{\rm in}$, where $w_{ij}$ is the link weight, ($z$-score), $s_i^{\rm in}$ denotes the in strength of journal $i$ (in number of citations), and $\alpha$ is corresponding to the teleportation probability. We have chosen the widely used $\alpha=0.15$ parameter value, however, the ordering of the journals according to the centralities was quite robust with respect to changes in $\alpha$. 

Based on the centralities a directed tree representing the backbone of the hierarchy is built from bottom up as described in ``Algorithm B'' in \citep{Tibely_plosone}. In the event that we cannot find a suitable ``parent'' for node $i$ according to the original rules, we chose the node with the highest accumulated $z$-score from all journals that have a higher centrality than $i$, (where the accumulation is running over the already found descendants of the given node). This ensures the emergence of a single connected component, since a single direct ancestor is assigned to every node (except for the root of the tree). This is followed by a final iteration over the nodes where we examine whether further `parents' have to be assigned or not. The criteria for accepting a node as the second, third, and so on, direct ancestor of journal $i$ are that it must have a higher centrality compared to $i$, and also the $z$-score has to be larger than the $z$-score between $i$ and its first direct ancestor. Note that the first parent is chosen based on aggregated $z$-score instead of the simple pairwise $z$-score, as explained in the work by \cite{Tibely_plosone}.

\subsection*{Nested hierarchy of scientific journals}

\begin{figure}[hbt]
\centerline{\includegraphics[width=\textwidth]{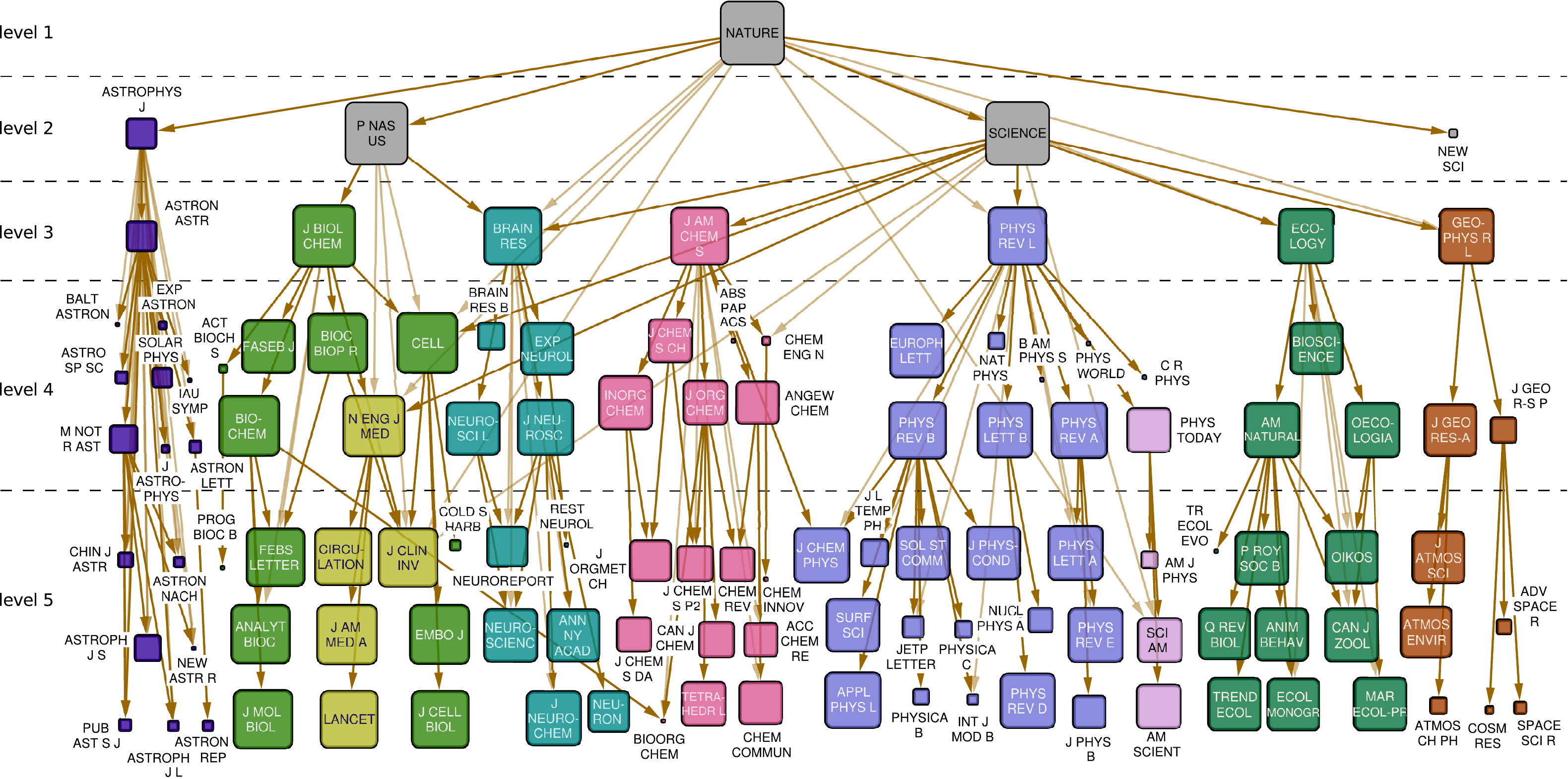}}
\caption{The top of the nested hierarchy between scientific journals. Due to the rapidly increasing number of nodes per level, the journals on levels four and five are organised into multiple rows. The size of the nodes indicates the total number of descendants (on a logarithmic scale). The journals positioned above level five with no out links shown, (e.g., Europhysics Letters or Bioscience), have descendants on levels that are out of the scope of the figure.}
\label{fig:algb_citations}
\end{figure}

In Fig.\ref{fig:algb_citations}. we show the top of the obtained nested hierarchy between the journals, with Nature appearing as the root, while PNAS, Science, New Scientist and The Astrophysical Journal form the second level. Several prominent field specific journals such as Physical Review Letters, Brain Research, Ecology and Journal of the American Chemical Society have both Nature and Science as direct ancestors. Interestingly, The Astrophysical Journal is a direct descendant only of Nature, and is not linked under Science, or PNAS. Nevertheless, it serves as a local root for a branch of astronomy related journals, in a similar fashion to Physical Review Letters, which can be regarded as the local root of physics journals, or Journal of the American Chemical Society, corresponding to the local root of chemical journals. The biological, medical and biochemical journals form a rather mingled branch under PNAS, with  Journal of Biological Chemistry as the local root and  New England Journal of Medicine corresponding to a sub-root for medical journals. However, Cell and  New England Journal of Medicine are direct descendants of Nature and Science as well. Interestingly, the brain- and neuroscience related journals form a rather well separated branch with Brain Research as the local root, linked directly under PNAS, Science and also under Nature. 

\section*{Comparing the hierarchies}
\label{Sect:compare}
Although the hierarchies presented in Fig.\ref{fig:GRC_hierarchy}.\ and Fig.\ref{fig:algb_citations}. show a great deal of similarity, some interesting differences can also be observed. The figures are showing the top of the corresponding hierarchies, and seemingly, a significant portion of the journals ranked high in the hierarchy are the same in both cases. However, the root of the hierarchies is different (Science in case of the flow hierarchy and Nature in case of the nested hierarchy), and also, the level-by-level comparison of Fig.\ref{fig:GRC_hierarchy}.\ and Fig.\ref{fig:algb_citations}.\ shows that a very high position in the flow hierarchy is not always accompanied by an outstanding position in the nested hierarchy, and vice versa. For example, Lancet and New England Journal of Medicine appear much higher in Fig.\ref{fig:GRC_hierarchy} compared to Fig.\ref{fig:algb_citations}, while  Geophysical Research Letters is just below Nature and Science in the nested hierarchy and is not even shown in the top of the flow hierarchy.

\begin{figure}[hbt]
\centerline{\includegraphics[width=0.6\textwidth]{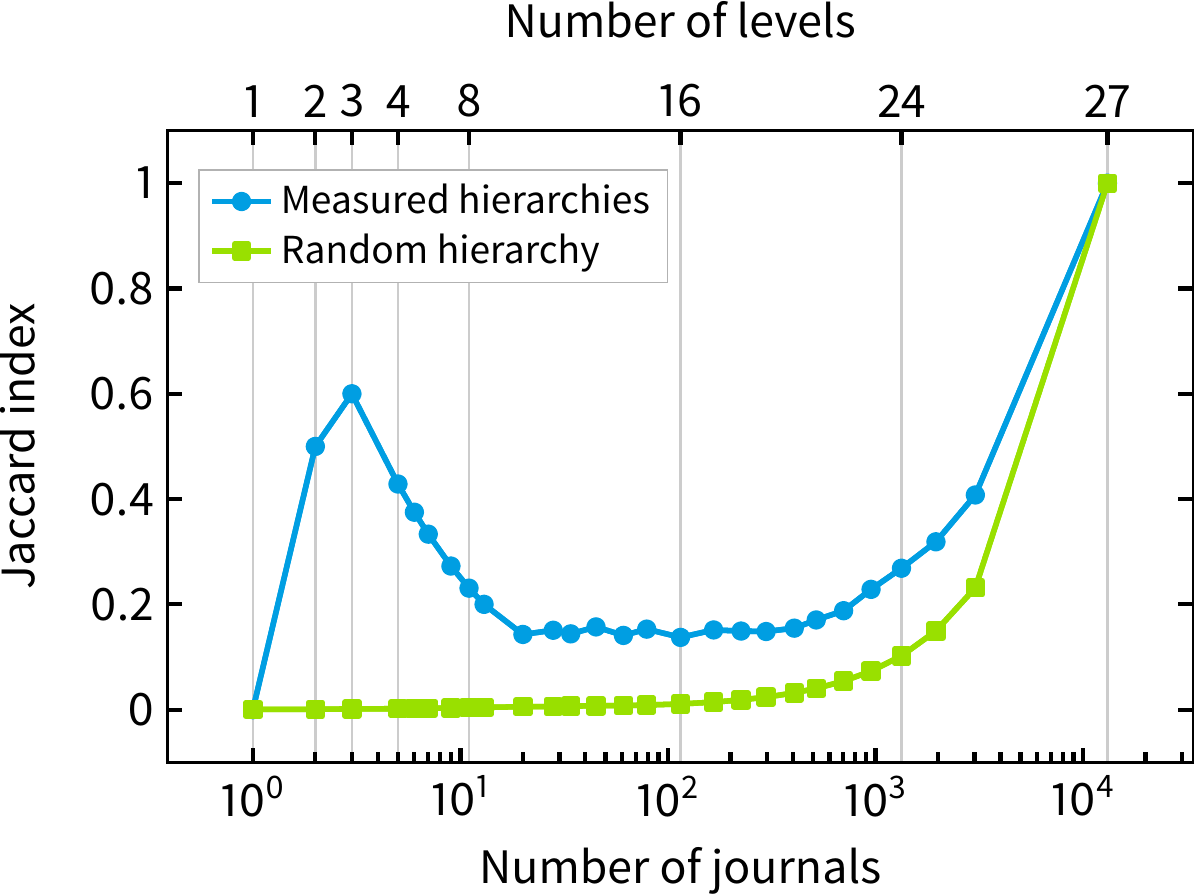}}
\caption{Comparison between the flow hierarchy and the nested hierarchy of scientific journals. The Jaccard similarity coefficient $J_f(\ell_f)$ of the aggregated sets of journals is plotted as a function of the number of accumulated journals from the top of the hierarchy to level $\ell_f$ in the flow hierarchy. Circles are corresponding to the similarity between the two hierarchies, while squares show the similarity between two random sets of journals of the corresponding size.}
\label{fig:Jaccard}
\end{figure}

To make the comparison between the two types of hierarchies more quantitative, we subsequently aggregated the levels in the hierarchies starting from the top, and calculated the Jaccard similarity coefficient between the resulting sets as a function of the level depth $\ell$. Thus, when $\ell=1$, we are actually comparing the roots, when $\ell=2$, the journals on the top two levels, and so on. However, since the total number of levels in the hierarchies are different, we refine the definition of the similarity coefficient by allowing different $\ell$ values in the two hierarchies, and always choosing the pairs of aggregated sets with the maximal relative overlap. Therefore, we actually have two similarity functions,
\begin{eqnarray}
J_{f}(\ell_f)&=&\max_{\ell_n}\frac{\left|S_f(\ell_f)\cap S_n(\ell_n)\right|}{\left|S_f(\ell_f)\cup S_n(\ell_n)\right|}, \label{eq:Jac_f}\\
 J_{n}(\ell_n)&=&\max_{\ell_f}\frac{\left|S_f(\ell_f)\cap S_n(\ell_n)\right|}{\left|S_f(\ell_f)\cup S_n(\ell_n)\right|}, 
\label{eq:Jac_n}
\end{eqnarray}
where $S_f(\ell_f)$ and $S_n(\ell_n)$ denote the set of aggregated journals from the root to level $\ell_f$ in the flow hierarchy and to level $\ell_n$ in the nested hierarchy, respectively. When evaluating $J_f(\ell_f)$ at a given level depth $\ell_f$ according to (\ref{eq:Jac_f}), the set of aggregated journals in the flow hierarchy, $S_f(\ell_f)$ is fixed, and we search for the most similar set of aggregated journals from the nested hierarchy by scanning over the entire range of possible $\ell_n$ values, and choose the one giving the maximal Jaccard similarity. Similarly, when calculating $J_n(\ell_n)$ according to (\ref{eq:Jac_n}), the set of aggregated journals taken from the nested hierarchy, $S_n(\ell_n)$ is fixed, and the set $S_f(\ell_f)$ yielding the maximal Jaccard similarity is chosen from the flow hierarchy.

In Fig.\ref{fig:Jaccard}.\ we show the result obtained for $J_f(\ell_f)$ as a function of the level depth $\ell_f$ in the flow hierarchy, (while the corresponding $J_n(\ell_n)$ plot for the nested hierarchy is given in Fig.S10. in the Supplementary Information S6.).  Beside $J_f(\ell_f)$,  in Fig.\ref{fig:Jaccard}.\ we also plotted the expected similarity between the aggregated sets of journals and a random set of journals of the same size. Since the roots of the hierarchies are different, the curves are starting from zero at $\ell_f=1$, and naturally, as we reach to the maximal level depth, the similarity is approaching to one, since all journals are included in the final aggregate. However, at the top levels below the root, a prominent increase can be observed in the $J_f(\ell_f)$, while the similarity between random sets of journals is increasing only very slowly in this region. Thus, the flow hierarchy and the nested hierarchy revealed by our methods show a significant similarity also from the quantitative point of view. This is also supported by the remarkably small $\tau=0.16$ generalised Kendall-tau distance obtained by treating the two hierarchies as partial orders, and applying a natural extension of the standard distance measure between total orders. The definition of the distance measure and the details of the calculation are given in the Supplementary Information S7.

Finally, our hierarchies can also be compared to traditional impact measures. According to the results detailed in the Supplementary Information S8, both the flow- and the nested hierarchy show moderate correlations with the impact factor, the Scimago Journal Rank and the closeness centrality of the journals in the aggregated citation network. Therefore, the general trends shown by the hierarchies are consistent with previously introduced, widely used impact measures. However, when looking into the details, they also provide an alternative point of view with important differences, circumventing large correlation values with the former, one dimensional characterisations of journal ranking.

%\section*{ Data and methods}

\section*{Discussion}
Ranking and comparing the importance, prestige and popularity of scientific journals is a far from trivial task with quite a few different available impact measures \citep{Garfield_impact,Garfield_impact_2,Braun_H_index,Egghe_g_index,Bergstrom_impact,Bollen_impact,Bollen_centralities,Leydesdorff_betweenness,Bollen_Plos,Franceschet_biblio,Franceschet_reasons,Glanzel_impact,Filippo_impact}. However, it seems that the overall impact of journals cannot be adequately characterised by a single one dimensional quality measure \citep{Bollen_Plos}. In this light, our results offer an informative overview on the ranking and the intricate relations between journals, where instead of e.g., simply ordering them according to a one dimensional parameter, we organise them into multiple hierarchies. 

First, we defined a flow hierarchy between the journals based on the $m$-reaching centrality in the citation network between the scientific papers. This structure organises the journals according to their potential for spreading new scientific ideas, with the most influential information spreaders sorted at the top of the hierarchy. In this respect Science turned out to be the root, followed by Nature and PNAS, and the top dozen levels of the hierarchy were dominated by multidisciplinary, biological, biochemical and medical journals. 

We also constructed a nested hierarchy between the journals  by generalising a recent tag hierarchy extraction algorithm. In this case the journals were organised into branches according to the major scientific fields, with a clear separation between unrelated fields, and relatively strong mixing and overlap between closely related fields. Mapping the different journals into well oriented knowledge domains is a complex problem on its own \citep{Paul_domains,Paul_jigsaw,Katy_knowledge_domains,Martin_impact,Katy_atlas}, especially from the point of view of multi- and interdisciplinary fields. Our nested hierarchy provides a  natural tool for the visualisation of the intricate nested and overlapping relations between scientific fields as well. An important feature is that the organisation of the branches roughly highlights the local hierarchy of the given field, with usually the most prominent journal in the field serving as the local root, and more specialised journals positioned at the bottom. Thus, zooming into a specific field for comparing and ranking the journals that publish in the given field becomes simple: we just have to select the corresponding branch in the nested hierarchy. 

 Another interesting perspective is that based on the position of a journal in the nested hierarchy we gain immediate information on its standing within its particular field. According to that we can select those journals with which we can make a fair comparison, and we can exclude journals in far away branches from any comparing study. Moreover, similarly to judging the position of a journal within its specific field (a local branch), we can also judge the standing of this sub-field in a larger scientific domain, (a main branch), and so on, and thereby, compare the ranking of the different scientific fields and sub-fields (each being composed of multiple journals). When zooming out completely to the overall hierarchy between the journals, Nature was observed to be in the very top position  with Science, PNAS, The Astrophysical Journal and  New Scientist formed the second level, and the field dependent branches starting at the third level.

The comparison between the two types of hierarchy reveals a strong similarity accompanied by significant differences. Basically, Science, Nature and PNAS provide the top three journals in both cases, and also, the top few hundred nodes in the hierarchy have a far larger overlap than expected at random. However, a closer level-by-level inspection showed that a very high position in, for example, the flow hierarchy does not guarantee a similarly outstanding ranking in the nested hierarchy, and vice versa. Both hierarchies showed moderate correlations with the impact factor, the Scimago Journal Rank and closeness centrality in the citation network. This supports our view that the hierarchical organisation of scientific journals provides an interesting alternative for the description of journal impact, which is consistent with the previously introduced measures at large, but in the mean time it shows important differences when examined in details.

In summary, the two hierarchies we constructed offer a compound view of the inter relations between scientific journals, and provide a higher dimensional characterisation of journal impact instead of ranking simply according to a one-dimensional parameter. Naturally, hierarchies between scientific journals can be defined in other ways too \citep{dominance_hierarchy}. For example, when building a flow hierarchy, the overall influence of journals could be measured alternatively with other quantities such as the wake-citation score \citep{wake_citation}, the PageRank or the Y-factor \citep{Bollen_impact}. In parallel, a nested hierarchy might also be constructed by suitably modifying a community finding algorithm producing inherently nested and overlapping communities such as the Infomap \citep{Martin_impact,Rosvall_memory,Martin_hier_infomap} or the clique percolation method \citep{CPM_nature}. Another interesting aspect we have not taken into account here is given by the time evolution of the citation network between the journals. Obviously, the ranking of the journals changes with time, and by treating all publications between 1975 and 2011 in a uniform framework we neglected this effect. However, the examination of the further possibilities for hierarchy construction and the study of the time evolution of the journal hierarchies is out of the scope of the present work, although it provides interesting directions for future research.

\section*{Additional information}
Supplementary information accompanies this paper.\\
The authors declare no competing financial interest.

\section*{Data availability}
The datasets analysed during the current study are available in the Web of Science repository, owned by Thomson Reuters, {\tt http://scientific.thomson.com/isi/ } but restrictions apply to the availability of these data, which were used under license from Thomson Reuters, and so are not publicly available. Data are however available from the authors upon reasonable request and permission of Thomson Reuters.

The downloading scripts used in the study are available in the Dataverse repository: {\tt http://dx.doi.org/10.7910/DVN/MCXTHF }

\section*{Acknowledgements}
The research was partially supported by the European Union and the European Social Fund through project FuturICT.hu (grant no: TAMOP-4.2.2.C-11/1/KONV-2012-0013), by the Hungarian National Science Fund (OTKA K105447) and by the EU FP7 ERC COLLMOT project (grant no: 227878). The funders had no role in study design, data collection and analysis, decision to publish or preparation of the manuscript.

%\bibliography{wos_refs_clean}

\newpage

\begin{center}
\Huge{\bf Supplementary Information}
\end{center}
\vspace{1cm}

\renewcommand{\thefigure}{S\arabic{figure}}
\renewcommand{\thetable}{S\arabic{table}}
\renewcommand{\theequation}{S\arabic{equation}}
\renewcommand{\thesection}{S\arabic{section}}
\setcounter{equation}{0}
\setcounter{figure}{0}

\section{The m-reach and the wake-citation-score}
The flow hierarchy we study is based on the $m$-reach defined for the journals in Eq.(1) in the main paper. However, there are also further other alternatives for measuring the impact of the journals or individual papers by taking into account deeper layers of other papers in the citation network. A prominent example is given by the wake-citation-score, introduced by %D.~F.~Klosik and S.~Bornholdt 
\cite{wake_citation}. The basic idea here is to calculate a weighted sum over the publications in the in-component of a given paper $i$ up to a certain maximal distance $d_{\rm max}$, where the weight of the papers is decreasing according to ${\alpha}^{d}$, where $\alpha\in[0,1]$ is a constant, and $d$ denotes the distance from $i$ in the citation network. In the work by \cite{wake_citation} the impact of the publications appearing in a given year are compared, thus, the wake-citation-score is normalised by the maximal score obtained in the given year.

Here we compare the $m$-reach with the wake-citation-score in case of scientific journals appearing in our database. However, in our case the normalisation by the largest wake-citation-score per year cannot be applied, as we are interested in the overall citation of the journals in the entire available time period. Therefore, the wake-citation-score was adapted to journals as follows. In order to have notation consistent with the main paper, we consider here a network where the links are pointing from a reference article to all papers citing it. First we carry out a weighted summation as
\begin{equation}
W(\mathcal{J})=\sum_{d=1}^{d_{\rm max}}\alpha^{d}\left|\left\lbrace
i \mid d_{\rm out}(j,i)=d, j\in\mathcal{J}\land i\notin \mathcal{J}\right\rbrace 
 \right|,
\end{equation}
where $W(\mathcal{J})$ is the (non-normalised) wake-citation-score of the journal $\mathcal{J}$ and $d_{\rm out}(j,i)$ denotes the out-distance from paper $j$ to $i$. When the quantity above has been evaluated for all journals in the data set, we simply normalise the results by the largest $W(\mathcal{J})$ obtained. 

\begin{figure}[hbt]
\centerline{\includegraphics[width=\textwidth]{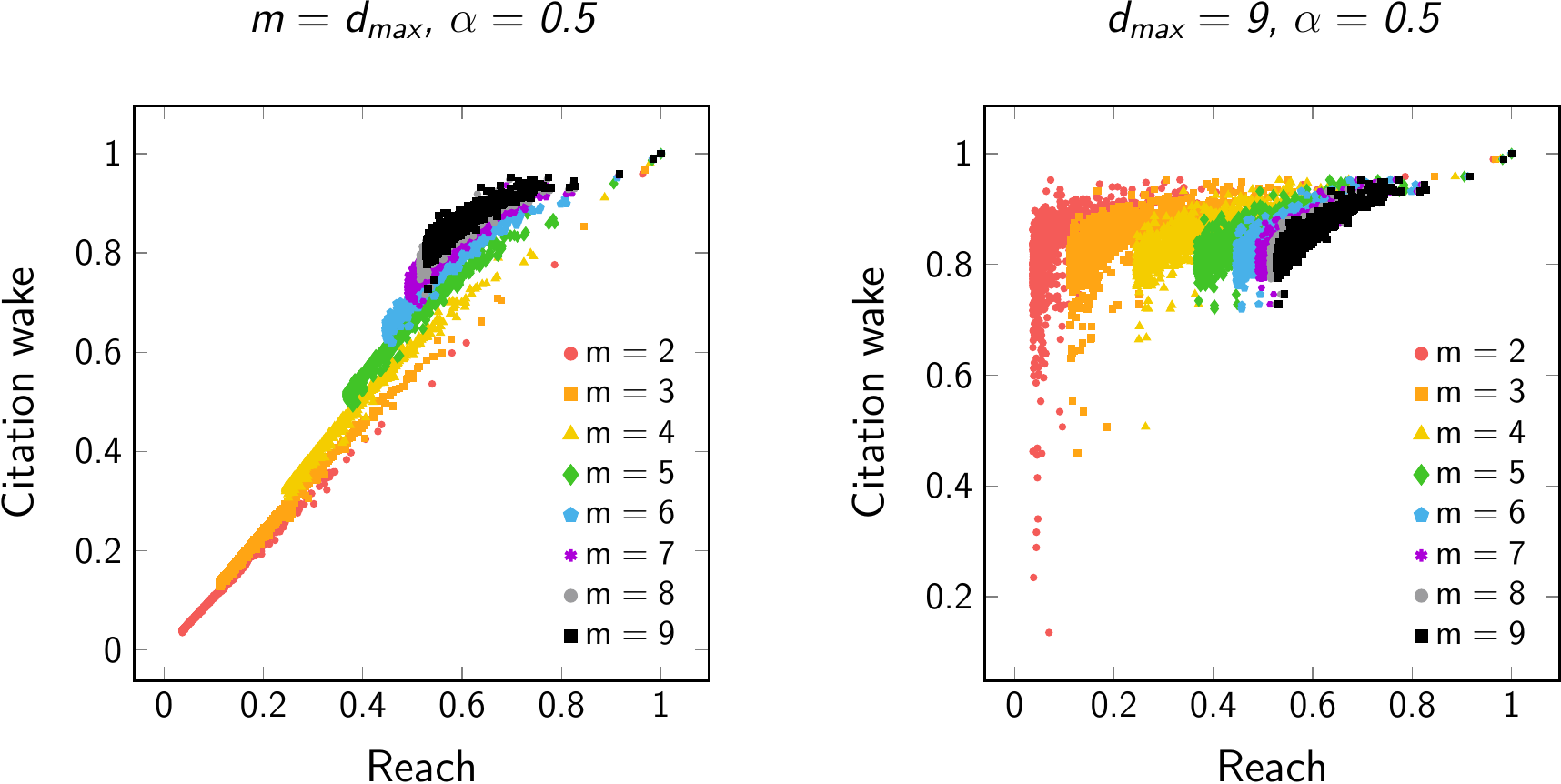}}
\caption{
a) The wake-citation-score of journals at $\alpha=0.5$ as a function of their $m$-reach, (where each data point is representing a different journal). The different colours are coding the results obtained for different $m$ values. The $d_{\rm max}$ in the calculation of the wake-citation-score was set to $d_{\rm max}=m$. b) The wake-citation-score of journals at $\alpha=0.5$ and $d_{\rm max}=9$ as a function of their $m$-reach.}
\label{fig:c_wake_m}
\end{figure}

In Fig.\ref{fig:c_wake_m}.\ we show the wake-citation-score of journals as a function of their $m$-reach. In order to ensure an $m$-reach value falling in the unit interval, the $m$-reach results were normalised by the largest $m$-reach obtained from the data set, (in a similar fashion to the wake-citation-score). In Fig.\ref{fig:c_wake_m}a we display the results obtained when $d_{\rm max}=m$ and $\alpha=0.5$, while Fig.\ref{fig:c_wake_m}b is showing the results for $d_{\rm max}=9$ and $\alpha=0.5$.
The obtained scatter plots suggest very strong correlations, especially in case of Fig.\ref{fig:c_wake_m}a. This is fully supported by the corresponding correlation values listed in Table \ref{table:c_wake_cors}.

\begin{figure}[hbt]
\centerline{\includegraphics[width=\textwidth]{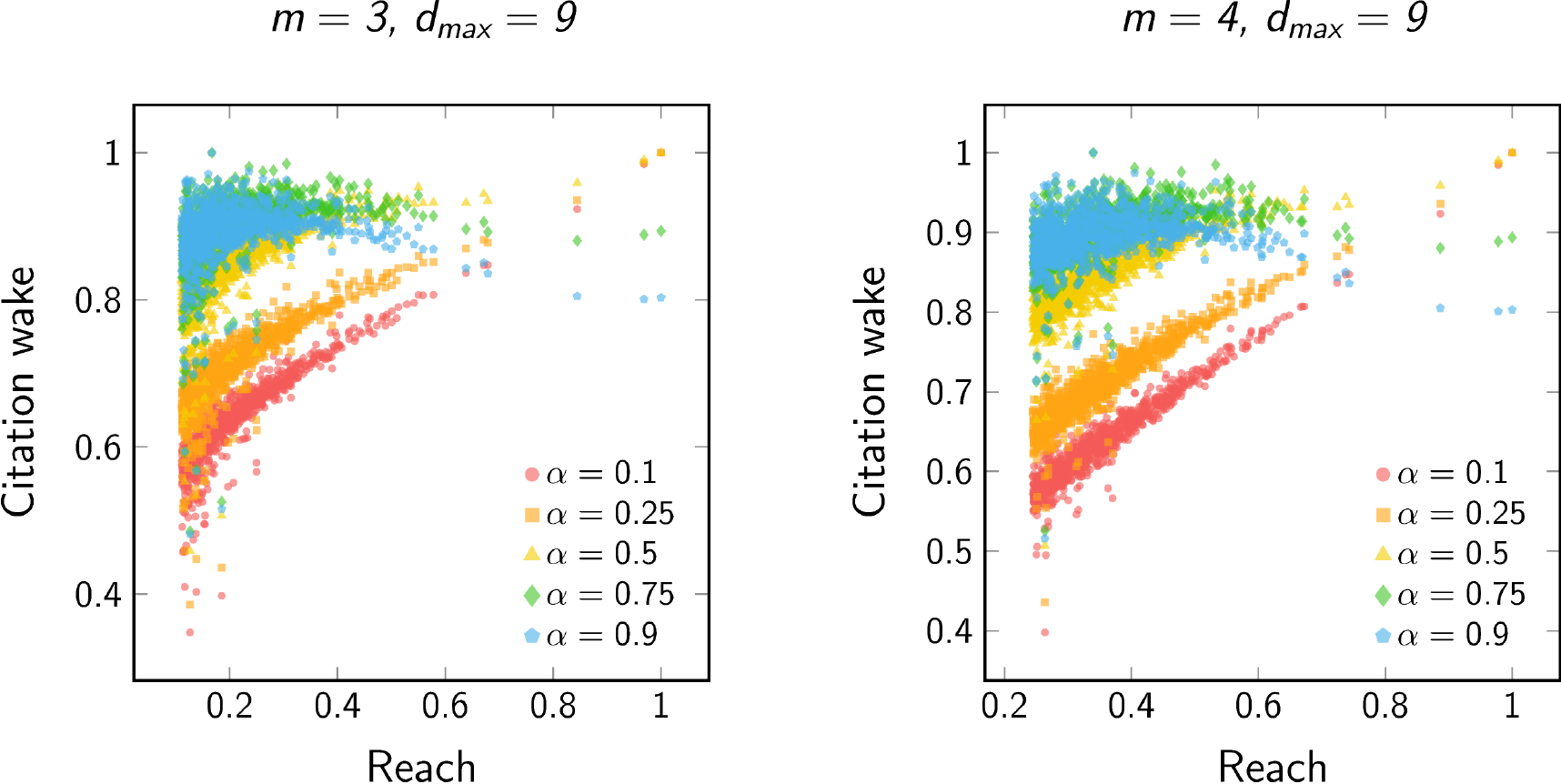}}
\caption{a) The wake-citation-score of journals at $d_{\rm max}=9$ as a function of their $m$-reach at $m=3$, (where each data point is representing a different journal). The different colours are coding the results obtained for different $\alpha$ values. b) The same plot as in case of a) when $m$ is set to $m=4$.}
\label{fig:c_wake_alfa}
\end{figure}

The dependence of the results on the parameter $\alpha$ is examined in Fig.\ref{fig:c_wake_alfa}.
Similarly to the previous case, a quite strong correlation can be observed between the wake-citation-score and the $m$-reach for the majority of the $\alpha$ values. The corresponding correlation values are given in Table \ref{table:c_wake_cors_2}.
\begin{table}[ht!] 
\centering
\tabcolsep=0.09cm
\footnotesize
\rowcolors{1}{lightgray}{white}
\begin{tabular}{|c|c|c|c|c|}
\hline
$m$ & $d_{\rm max}$ & $C_{\rm Pearson}$ & $C_{\rm Spearman}$ & $C_{\rm Kendall}$ \\ 
1 & 1 & 1.000 & 1.000 & 1.000 \\
2 & 2 & 0.999 & 1.000 & 0.994 \\
3 & 3 & 0.999 & 1.000 & 0.993 \\
4 & 4 & 0.999 & 1.000 & 0.993 \\
5 & 5 & 0.998 & 1.000 & 0.993 \\
6 & 6 & 0.997 & 1.000 & 0.991 \\
7 & 7 & 0.996 & 1.000 & 0.989 \\
8 & 8 & 0.995 & 1.000 & 0.985 \\
9 & 9 & 0.993 & 0.999 & 0.982 \\
\hline
\end{tabular}
\rowcolors{1}{lightgray}{white}
\begin{tabular}{|c|c|c|c|c|}
\hline
$m$ & $d_{\rm max}$ & $C_{\rm Pearson}$ & $C_{\rm Spearman}$ & $C_{\rm Kendall}$ \\ 
1 & 9 & 0.332 & 0.884 & 0.712\\
2 & 9 & 0.487 & 0.932 & 0.786\\
3 & 9 & 0.657 & 0.961 & 0.842\\
4 & 9 & 0.788 & 0.978 & 0.884\\
5 & 9 & 0.883 & 0.988 & 0.916\\
6 & 9 & 0.943 & 0.994 & 0.940\\
7 & 9 & 0.974 & 0.997 & 0.959\\
8 & 9 & 0.988 & 0.999 & 0.973\\
9 & 9 & 0.993 & 0.999 & 0.982\\
\hline
\end{tabular}
\caption{The correlation between the $m$-reach and the wake-citation score for the data shown in Fig.\ref{fig:c_wake_m}a (left) and for the data shown in Fig.\ref{fig:c_wake_m}b (right). (The $\alpha$ parameter in the calculation of the wake-citation-score was set to $\alpha=1/2$.) The first two columns are listing $m$ and $d_{\rm max}$, the 3$^{\rm rd}$ column provides the Pearson correlation, the 4$^{\rm th}$ column gives the Spearman's rank correlation coefficient, while the 5$^{\rm th}$ column is containing the Kendall rank correlation coefficient.}
\label{table:c_wake_cors}
\end{table}

\begin{table}[ht!] 
\centering
\tabcolsep=0.09cm
\footnotesize
\rowcolors{1}{lightgray}{white}
\begin{tabular}{|c|c|c|c|c|c|}
\hline
$m$ & $d_{\rm max}$ & $\alpha$ & $C_{\rm Pearson}$ & $C_{\rm Spearman}$ & $C_{\rm Kendall}$ \\ 
3 & 9 & 0.10 & 0.719 & 0.966 & 0.856\\
3 & 9 & 0.25 & 0.696 & 0.964 & 0.851\\
3 & 9 & 0.50 & 0.657 & 0.961 & 0.842\\
3 & 9 & 0.75 & 0.619 & 0.957 & 0.832\\
3 & 9 & 0.90 & 0.596 & 0.953 & 0.823\\
\hline
\end{tabular}

\rowcolors{1}{lightgray}{white}
\begin{tabular}{|c|c|c|c|c|c|}
\hline
$m$ & $d_{\rm max}$ & $\alpha$ & $C_{\rm Pearson}$ & $C_{\rm Spearman}$ & $C_{\rm Kendall}$ \\ 
4 & 9 & 0.10 & 0.840 & 0.982 & 0.899\\
4 & 9 & 0.25 & 0.821 & 0.980 & 0.894\\
4 & 9 & 0.50 & 0.788 & 0.978 & 0.884\\
4 & 9 & 0.75 & 0.754 & 0.974 & 0.873\\
4 & 9 & 0.90 & 0.734 & 0.971 & 0.864\\
\hline
\end{tabular}
\caption{The correlation between the $m$-reach and the wake-citation score for the data shown in Fig.\ref{fig:c_wake_alfa}a (top) and for the data shown in Fig.\ref{fig:c_wake_alfa}b (bottom). The first two columns are giving $m$ and $d_{\rm max}$, and the 3$^{\rm rd}$ column is showing the $\alpha$ parameter. The corresponding Pearson correlation is given in the 4$^{\rm th}$ column, followed by the Spearman's rank correlation coefficient in the 5$^{\rm th}$ column, and the Kendall rank correlation coefficient in the 6$^{\rm th}$ column.}
\label{table:c_wake_cors_2}
\end{table}

\section{Setting the parameter m in the flow hierarchy analysis}
\begin{figure}[hbt]
\centerline{\includegraphics[width=0.6\textwidth]{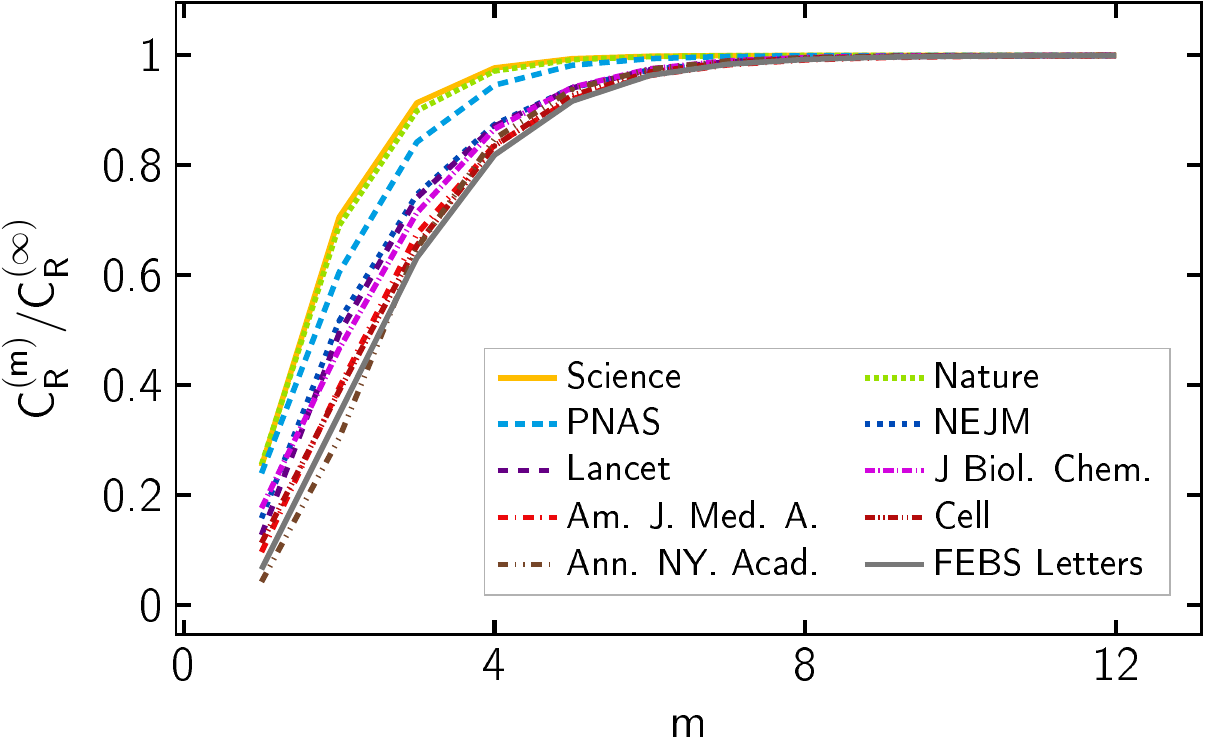}}
\caption{
The size of the $m$-reachable set, denoted by $C_m(\mathcal{J})$, divided by the size of the reachable set of papers at unlimited path length, as a function of $m$ for the top 10 journals according to the flow hierarchy.
}
\label{fig:saturation}
\end{figure}

The maximal allowed path length in the calculation of the $m$-reaching centrality, denoted by $m$, is an important parameter of our approach for the analysis of the flow hierarchy between scientific journals. Since the structure of the citation network is very far from a crystal lattice or a regular tree, and it also has a relatively large link density, the small world effect is expected to take place: the average distance is low between pairs of papers that can be reached from one to the other following the citations. Thus, the number of reachable articles from a given paper or a given journal saturates rather fast as a function of $m$. This effect is shown in Fig.\ref{fig:saturation}., where we plot the $C_m(\mathcal{J})$, corresponding to the size of the $m$-reachable set of papers (defined in Eq.(1) in the main paper), divided by the size of the reachable set of papers at unlimited path length $m\rightarrow\infty$, as a function of $m$ for the top 10 journals. According to the curves, the reach of Science and Nature saturates already at $m=4$, the reach of PNAS around $m=5$, while for the rest of the journals in the figure, the saturation occurs at higher $m$ values. 

The two main reasons for the saturation effect are the exponential increase of the number reachable papers as a function of $m$, and the finite system size. Since the saturation effect takes place at different $m$ values for the different journals, in order to provide a fair comparison between their influence based on the ``information spreading ability'', we should take an $m$ value below the saturation of all journals, that is an $m$ value below $m<4$. When $m>4$, the already saturated journals have a disadvantage, as their $m$-reach is already starting to be affected by the finite system size, while the not yet saturated journals do not suffer from this problem. 

Keeping $m$ smaller than $m=4$ is also consistent with the general intuition about the spread of information on the citation network: a direct citation is usually corresponding to a strong interrelation between the two papers, which are likely to be focused on the same field. However, as we increase the distance between the papers in the citation network, the relatedness between them usually drops, e.g., a pair of papers 4 citation steps away from each other can very easily belong to absolutely different fields.

Based on the above, we have chosen to set $m$ to $m=3$ in the flow hierarchy analysis outlined in the main paper. According to Fig.\ref{fig:saturation}., on the one hand this way we avoid the saturation effect present at $m\geq 4$ values. On the other hand, we also allow multiple steps in the information spread over the system, with a limited path length where we can still assume at least a weak relatedness between the papers at the opposite end of the citation path. A further advantage of this choice is that variance of the $C_m(\mathcal{J})$ values is significantly larger at $m=3$ compared to e.g., $m=5$, thus, providing a ranking between the journals based on $C_m(\mathcal{J})$ is much more robust at $m=3$.

\section{ Flow hierarchy at different $m$ values }
\begin{figure}[hbt]
\centerline{\includegraphics[width=\textwidth]{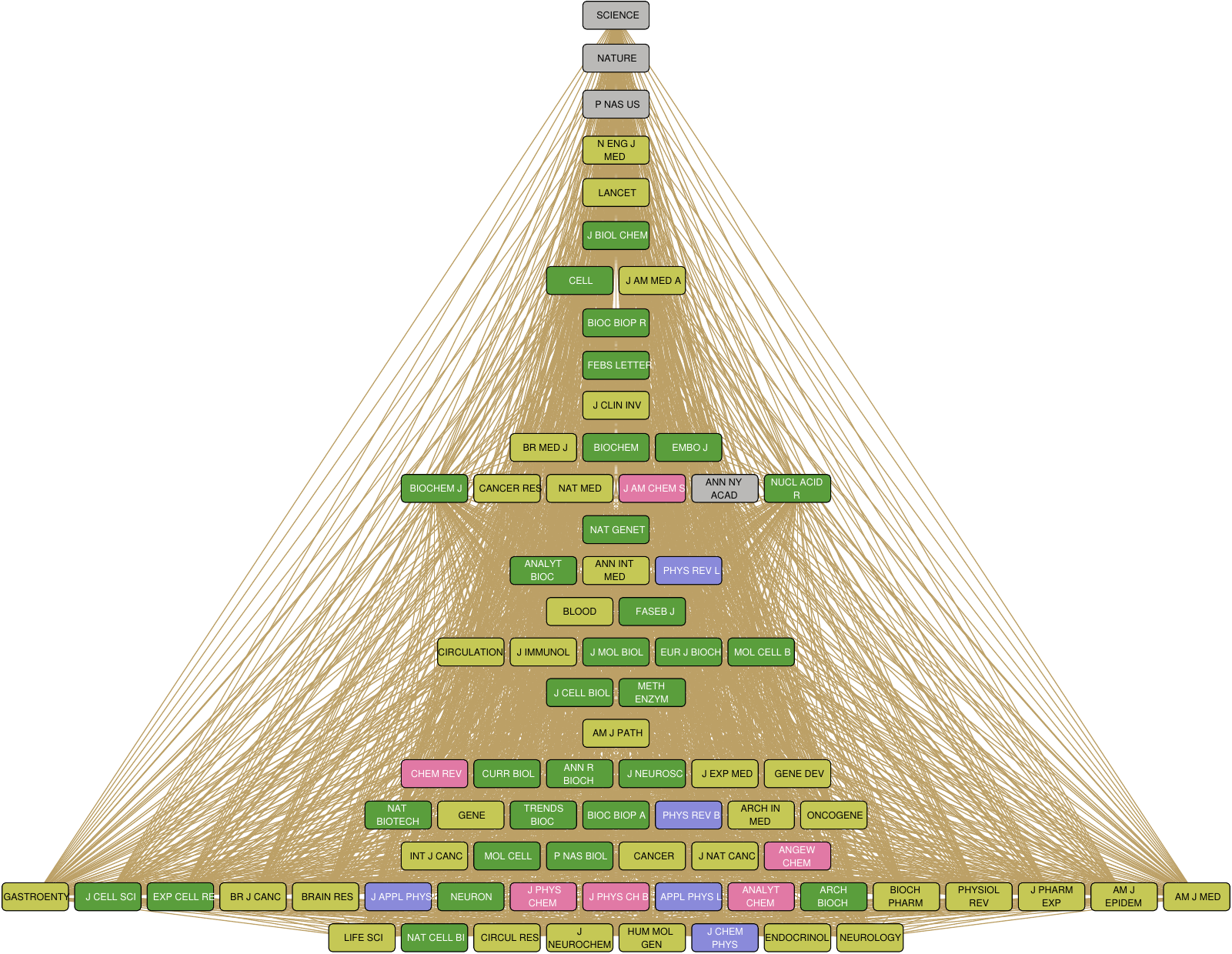}}
\caption{
The top of the flow hierarchy when $m$ is set to $m=2$, and the standard variation of $C_m$ withing a level is at most $0.13\cdot \sigma(C_m)$, where $\sigma(C_m)$ is corresponding to the standard variation of $C_m$ over all journals.
}
\label{fig:flow_m2}
\end{figure}

In order to fully complete the analysis of the effect of the choice of the parameter $m$ on the flow hierarchy, in this Section we show results obtained when $m$ is set to lower values compared to the optimal $m=3$ case. In Fig.\ref{fig:flow_m2}.\ we present the top of the hierarchy at $m=2$. Apparently, the very peak of the hierarchy is looking very similar to the $m=3$ case, given in Fig.2. in the main paper. I.e., Science is on the top, followed by Nature, with PNAS coming 3$^{\rm rd}$, while the New England Journal of Medicine and Lancet are just below the three major interdisciplinary journals. However, as we go deeper down the hierarchy levels, the difference between the two hierarchies becomes visible. E.g., the top chemical and physical journals gain a relatively higher position in Fig.\ref{fig:flow_m2}.\ compared to Fig.2.\ in the main paper.

\begin{figure}[hbt]
\centerline{\includegraphics[width=\textwidth]{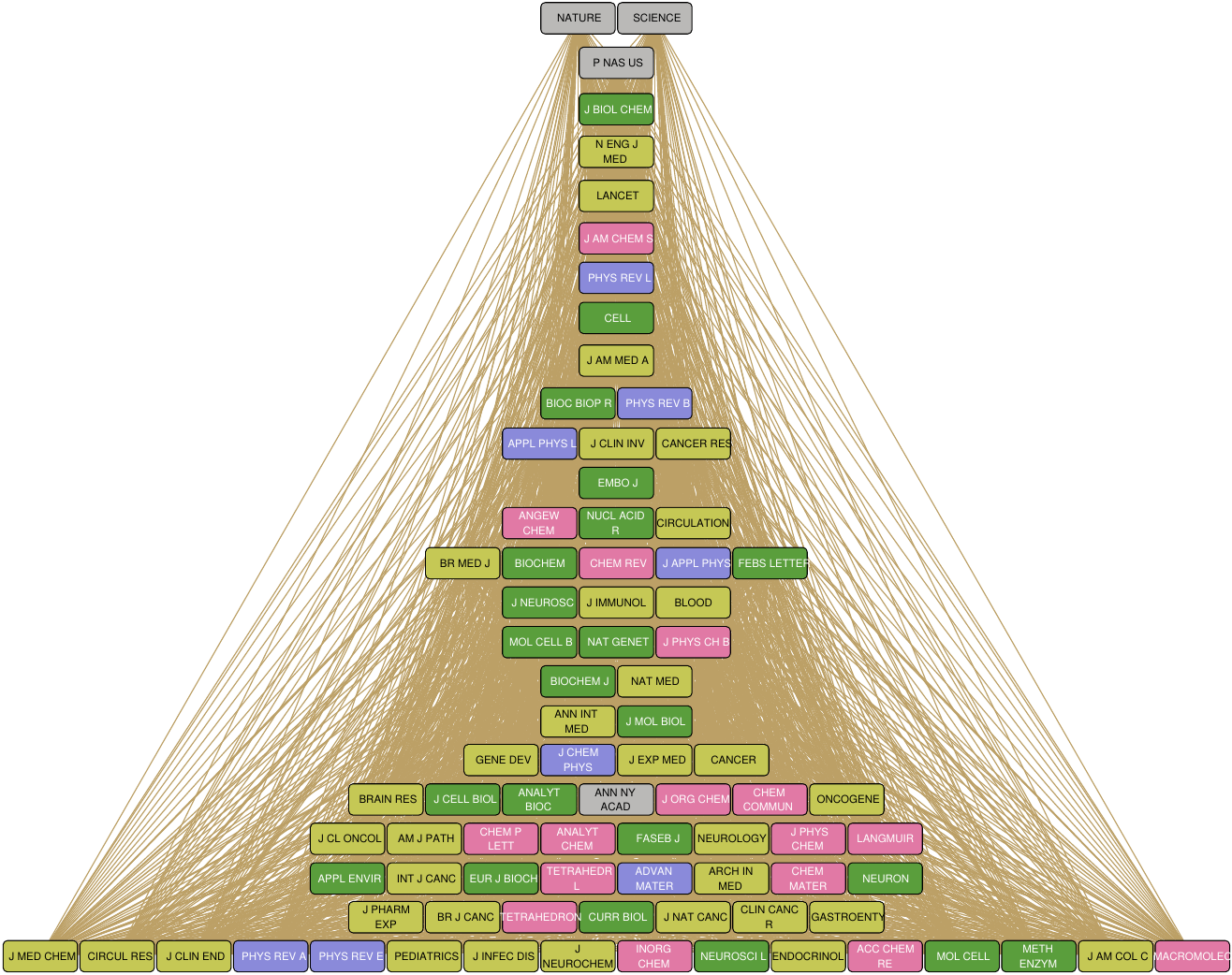}}
\caption{
The top of the flow hierarchy when $m$ is set to $m=1$, and the standard variation of $C_m$ withing a level is at most $0.13\cdot \sigma(C_m)$, where $\sigma(C_m)$ is corresponding to the standard variation of $C_m$ over all journals.
}
\label{fig:flow_m1}
\end{figure}

When switching to $m=1$, the discrepancies become much stronger, as shown in Fig.\ref{fig:flow_m1}. In this case the top position is shared by Nature and Science, while PNAS is placed on the $2^{\rm nd}$ level, followed by the Journal of Biological Chemistry on the $3^{\rm rd}$ level. Although the New England Journal of Medicine and Lancet are still just below the peak on the $4^{\rm th}$ and $5^{\rm th}$ levels, respectively, the Journal of the American Chemical Society and Physical Review Letters are overcoming the rest of the biological, biochemical and medical journals. In parallel, the fraction of physical and chemical journals is significantly higher in the overall picture of the top the hierarchy compared to the $m=3$ case shown in Fig.2. in the main paper.

We can also calculate the correlation between $C_m$ at the optimal $m=3$ setting, and the $C_m$ obtained for lower $m$ values. The Pearson correlation coefficient between the results obtained at $m=3$ and $m=2$ is $C_{\rm Pearson}=0.922$, while the Spearman's rank correlation for the same data is $C_{\rm Spearman}=0.991$, indicating a quite high similarity between the two rankings. However, when lowering $m$ to $m=1$, the corresponding correlation coefficients are decreasing to $C_{\rm Pearson}=0.724$ and $C_{\rm Spearman}=0.955$. Based on the above, the structure of the flow hierarchy is moderately robust against changes in the parameter $m$. I.e., if decreasing the length of the maximally allowed citation chains by one, the overall picture of the top of the hierarchy remains the same, with some differences becoming apparent in a level by level comparison. However, if we apply a more drastic change in $m$, the differences are becoming stronger, affecting also the very top ranks of the hierarchy.

\section{Aggregated citation network}
An alternative option for analysing the hierarchical relations of scientific journals based on publication data is to first construct a citation network between journals instead of individual papers, and in the next step apply the hierarchy related methods on the level of this aggregated network. The weights of the directed links between the journals in this framework are corresponding to the accumulated number of papers appearing in the ``target'' journal citing at least one paper appearing in the ``source'' journal. The advantage of this approach is that journals are represented by single nodes in the obtained network instead of groups of nodes as in case of the citation network between papers. However, a considerable drawback is that scientific citation networks have a memory \citep{Rosvall_memory}: e.g., a paper citing mostly biological articles is likely to be cited mainly by biological papers as well. When calculating  e.g., the reaching centrality of a journal in the aggregated network we neglect this memory effect, and thus, the result can show large deviations compared to the value obtained in the original citation network between papers. This effect is also very closely related to the distortions that can be caused by time aggregation in temporal networks, as  pointed out by \cite{Ingo_aggregate}. Nevertheless, it is still worth analysing the hierarchical properties of the aggregated citation network between journals for comparison with the results shown in Fig.2.\ in the main paper, with bearing in mind that the reaching centralities obtained here are somewhat distorted.

In order to concentrate only on the highly significant connections between the journals, we applied a weight threshold, $w^*$, taking into account only the links with a weight $w>w^*$. The weights of the links are distributed according to a power-law, inferring no plausible threshold by simply studying their distribution. Therefore, the final threshold was chosen so that the extent of hierarchy in the resulting network be maximal. A natural measure for the hierarchy is given by the Global Reaching Centrality (GRC) \citep{Enys_hierarchy}, reflecting the inhomogeneity of the reach of the individual nodes. The mathematical definition of the GRC is given by
\begin{equation}
  GRC_m=\frac{1}{N-1}\sum_{i}\left[\max\left\lbrace C_m(i)\right\rbrace-C_m(i)\right],
\label{eq:GRC}
\end{equation}
where $\max\left\lbrace C^{(m)}_R(i)\right\rbrace$ is the largest centrality in the network and $N$ is the number of nodes. 

\begin{figure}[hbt]
\centerline{\includegraphics[width=0.6\textwidth]{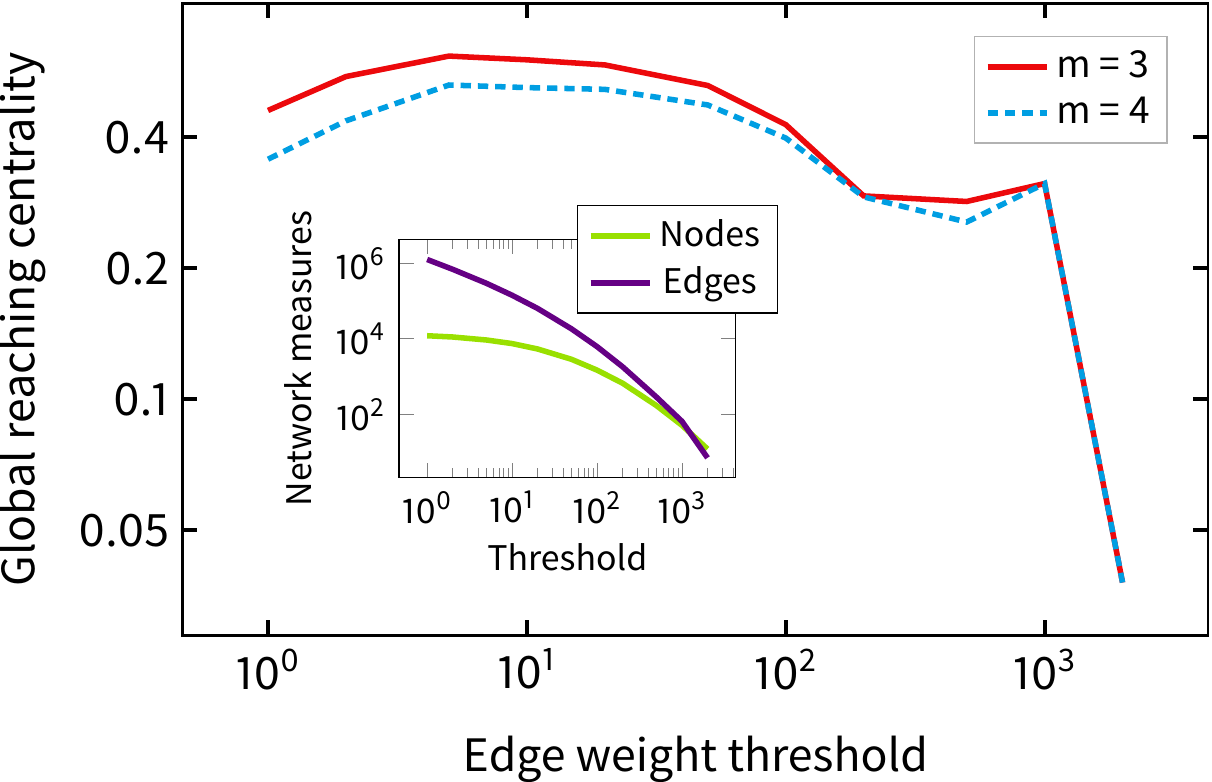}}
\caption{The $m$-reaching centrality in networks obtained by different edge-weight cutoffs. After the filtering of the edges, only those with a weight larger than $K_w\langle w\rangle$ are kept together with the corresponding nodes. The inset shows the number of nodes and edges as a function of the cutoff threshold.
}
\label{fig:GRC_threshold}
\end{figure}

Based on the above, we rejected links with a weight lower than $w^*=K_w\langle w\rangle$ where $\langle w\rangle$ denotes the average weight, allowing for different values of $K_w$. Afterwards, the centralisation of the m-reach centrality was calculated according to (\ref{eq:GRC}). In Fig.\ref{fig:GRC_threshold}.\ we show the obtained GRC$_m$ for $m=3$ as a function of $K_w$, with a clear global maximum at $K_w=10$. Thus, we applied this value in the investigation of the flow hierarchy at the level of the aggregated network between journals.

\begin{figure}[hbt]
\centerline{\includegraphics[width=\textwidth]{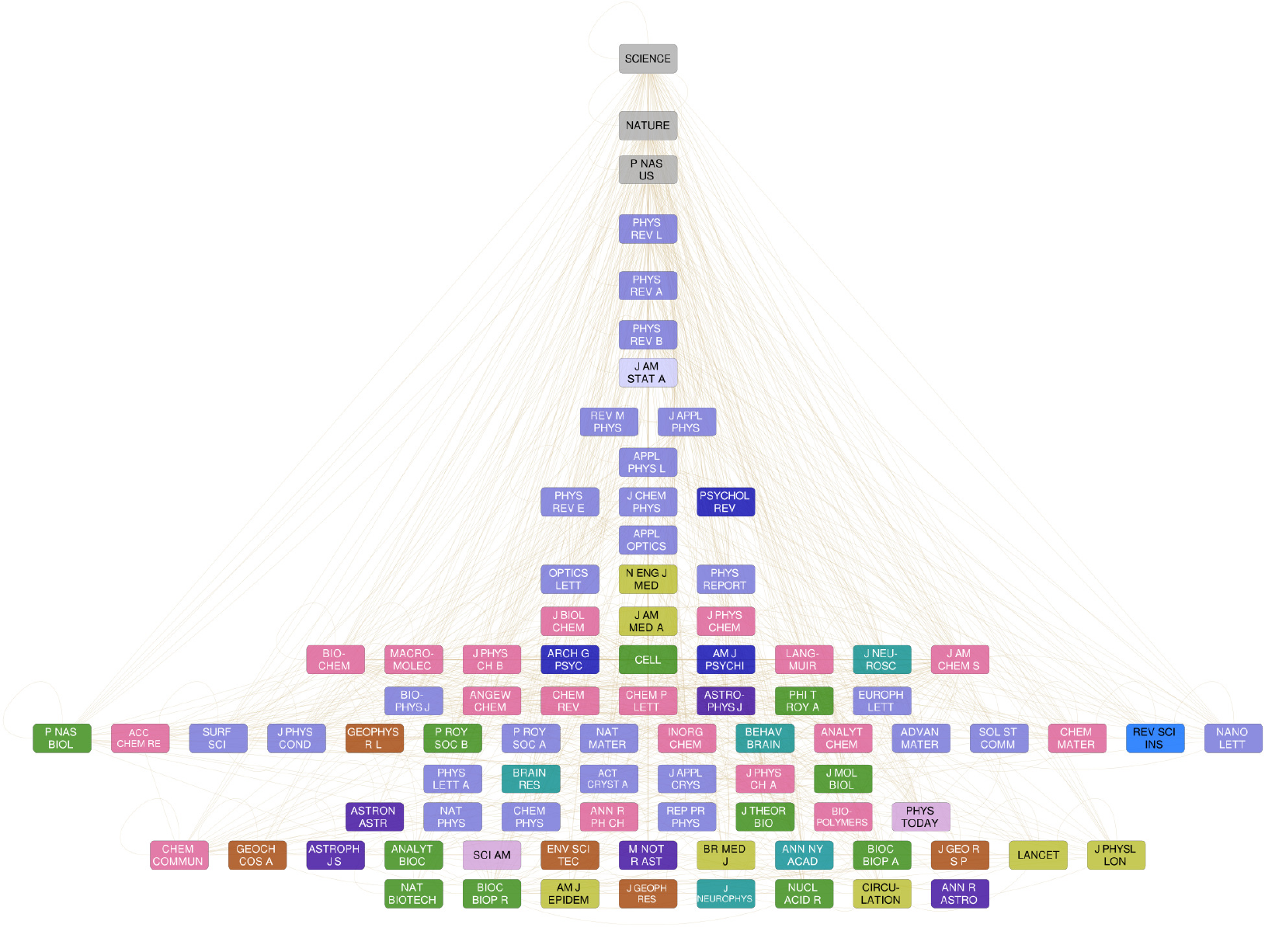}}
\caption{
Top of the flow hierarchy according to the $m$-reaching centrality $C_m$ at $m=3$, based on the aggregated network between the journals. The standard deviation of $C_m$ within the individual hierarchy levels is at most $0.13 \cdot \sigma(C_m)$, where $\sigma(C_m)$ denotes the standard deviation of $C_m$ over all journals.
}
\label{fig:GRC_aggr}
\end{figure}

In Fig.\ref{fig:GRC_aggr}.\ we show the top journals according to the reaching centrality within $m=3$ steps. Similarly to Fig.2.\ in the main paper, the hierarchy levels are obtained by aggregating the journals into subsets with a standard deviation of $C_m$ smaller than $0.13 \cdot \sigma(C_m)$, where $\sigma(C_m)$ denotes the standard deviation of $C_m$ over all journals. According to Fig.\ref{fig:GRC_aggr}., Science is the most influential journal according to the flow hierarchy, followed by Nature, with PNAS and Physical Review Letters are forming the 3$^{\rm d}$ level. Interestingly, physical journals dominate the next few levels, with Physical Review A on the 4$^{\rm th}$ level, Journal of Applied Physics on the 5$^{\rm th}$ level and Physical Review B and Physical Review E providing the 6$^{\rm th}$ level, followed by Applied Physics Letters on the 7$^{\rm th}$ level. 

This tendency is rather different from the results obtained from the citation network between individual papers (Fig.2.\ in the main paper), where medical, biological and biochemical journals occupied the top of the hierarchy. A plausible explanation is that when collapsing all the papers appearing in a given journal into a single node, we loose the information about the number of publications appearing in the journal. Since medical, biological and biochemical papers tend to cite mainly within these three fields, the reach of related journals is strongly reduced when switching from the network on the level of publications to the network between journals: The very high publication rate of these journals provides a high reach in the original network between papers, while the collapse of the vast number of papers appearing in these journals onto a single node in the aggregated network cancels out this effect. In contrast, papers appearing in physical journals have a somewhat larger likelihood for citing publications from other fields, thus, the aggregation of the papers into nodes representing journals does not have such a drastic effect on the reach. 

\section{ Changing the width of the levels in the flow hierarchy }

The levels in the flow hierarchy are defined based on the standard variation of the $m$-reach, i.e., $\sigma(C_m)$ within a level cannot exceed a fixed fraction of $\sigma(C_m)$ calculated over the entire set of journals. For simplicity, let us denote this parameter by $\omega$, thus, the maximal variance allowed for journals falling in the same level in the hierarchy is given by $\omega \sigma(C_m)$, where $\omega\in[0,1]$. Naturally, the larger $\omega$ we choose, the more journals we find in a hierarchy level on average, thus, the choice of this parameter has an effect on the overall shape of the hierarchy we obtain. 

\begin{figure}[hbt]
\centerline{\includegraphics[width=0.9\textwidth]{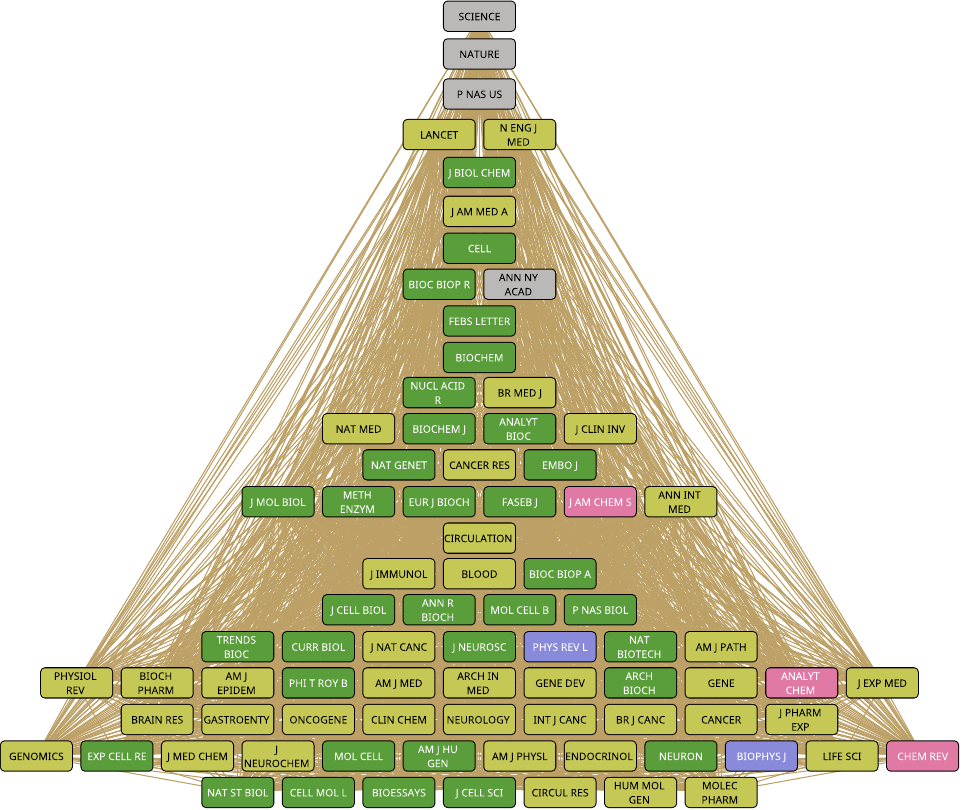}}
\caption{
The top of the flow hierarchy at $m=3$ when the standard variation of $C_m$ within a level is at most $0.08\cdot \sigma(C_m)$, where $\sigma(C_m)$ denotes the standard variation of $C_m$ over all journals.}
\label{fig:flow_sigma_008}
\end{figure}

\begin{figure}[hbt]
\centerline{\includegraphics[width=0.82\textwidth]{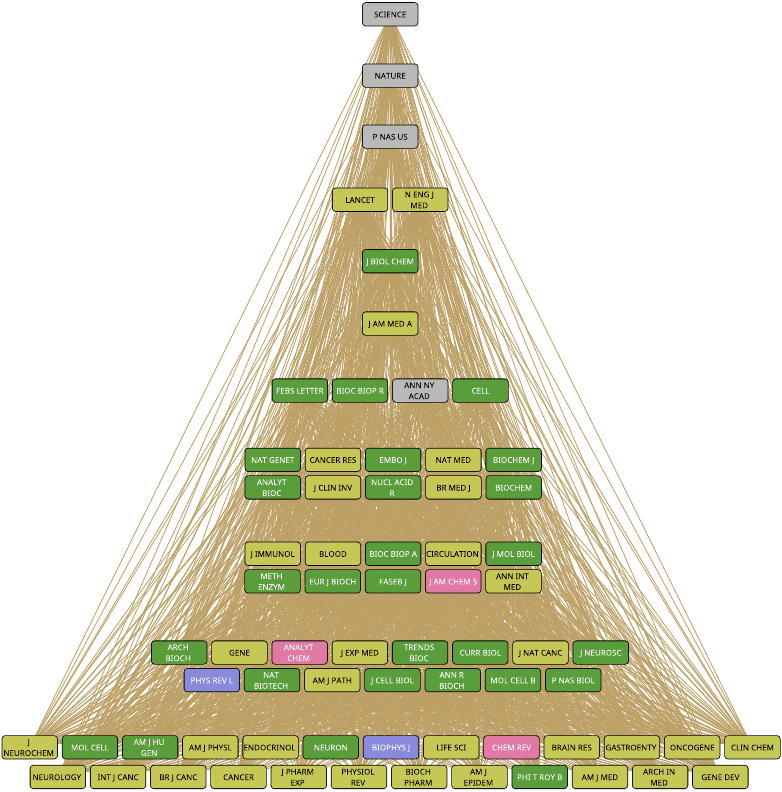}}
\caption{The top of the flow hierarchy at $m=3$ when the standard variation of $C_m$ within a level is at most $0.2\cdot \sigma(C_m)$, where $\sigma(C_m)$ denotes the standard variation of $C_m$ over all journals. Due to the rapidly increasing number of nodes per level, the journals on level eight and below are organised into double rows.}
\label{fig:flow_sigma_2}
\end{figure}

In the main paper we have shown the results for $\omega=0.13$. Here we provide visualisations of the top of the flow hierarchy at different $\omega$ values as well. In Fig.\ref{fig:flow_sigma_008}.\ we present the results for $\omega=0.08$, while in Fig.\ref{fig:flow_sigma_2}.\ we display the flow hierarchy at $\omega=0.2$.
\newpage

Although the width of the levels is sensitive to the choice of $\omega$ in general, the very top of the hierarchy is very robust. I.e., the first 6 levels are exactly the same in Figs.\ref{fig:flow_sigma_008}-\ref{fig:flow_sigma_2}.\ and Fig.2. in the main paper. When we go deeper in the hierarchy, naturally, the actual choice of $\omega$ is starting to make a difference, resulting in a narrow, steep overall shape at $\omega=0.08$, and a more wide, gradual overall shape at $\omega=0.2$.

A very useful property of journal hierarchies is that they provide an instant and simple visualisation of the journal rankings. However, when $\omega$ is low, the widening of the levels is slow as we go deeper in the hierarchy from the root, and therefore, the total number of journals that can be fitted in a picture of the top of the hierarchy is relatively low: When reaching the maximum level depth allowed by the height of the picture, (and the condition that the journal names should be still readable), the bottom levels are still rather narrow, as in case of \ref{fig:flow_sigma_008}. 

In contrast, when $\omega$ is large, the levels become wide fast as a function of the level depth. Thus, in this case we reach the maximally allowed width of the picture at a relatively low level depth from the root, and again, the total number of journals appearing in the visualisation is relatively low, in a similar fashion to \ref{fig:flow_sigma_2}. Based on the above, our choice of $\omega=0.13$ used in the main paper is corresponding to an optimal choice, allowing a relatively large number of journals in the visualisation of the top of the hierarchy.

\section{Jaccard similarity}
\begin{figure}[hbt]
\centerline{\includegraphics[width=0.6\textwidth]{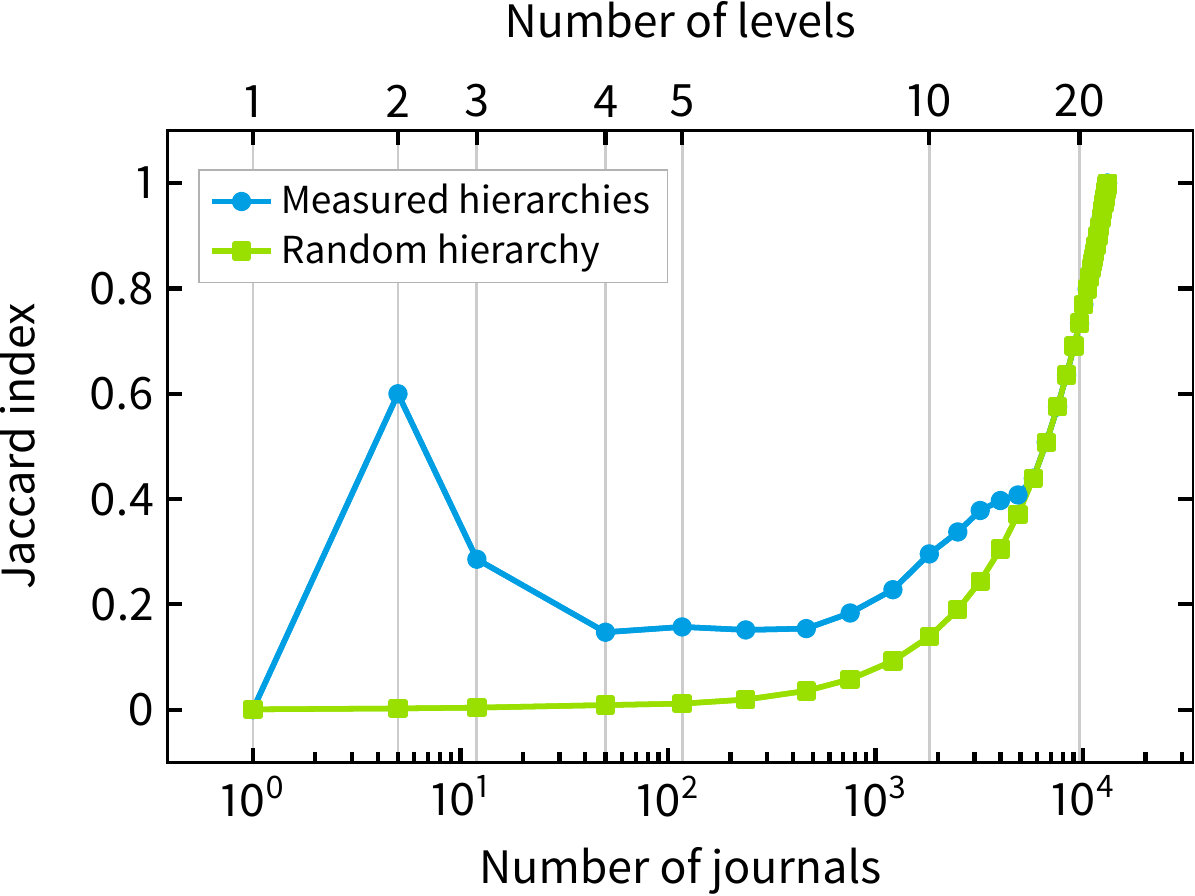}}
\caption{The Jaccard similarity $J_n(\ell_n)$ defined in Eq.(3) in the main paper, as a function of the level depth $\ell_n$ in the nested hierarchy.
}
\label{fig:Jac_nested}
\end{figure}

In the main paper we are comparing the flow- and the nested hierarchies based on the Jaccard similarity between the sets of aggregated journals from the root down to a certain level $\ell$. However, since the number of levels in the flow- and the nested hierarchy are different, we need to introduce actually a separate similarity measure for each hierarchy, as given in Eqs.(2-3) in the main paper. In Fig.4. in the main paper we displayed $J_f(\ell_f)$ for the flow hierarchy, here in Fig.\ref{fig:Jac_nested}.\ we show the corresponding $J_n(\ell_n)$ for the nested hierarchy. The behaviour is very similar to that of the $J_f(\ell_f)$ measure.

\section{Comparing the hierarchies by the Kendall-tau distance}
\label{sect:Kendall}
The studied flow- and the nested hierarchies can be compared also according to partial order distance measures. The basic idea is to first map the given hierarchy onto a partial order, given by a domain of candidates ${\mathcal C}$ and a relation ${\kappa}$ obeying the following conditions:
\begin{itemize}
\item $\kappa$ is irreflexive, i.e., $\forall x\in {\mathcal C}$ $x\not\prec_{\kappa} x$,
\item $\kappa$ is asymmetric, i.e., $x\prec_{\kappa} y$ $\Rightarrow$ $y\not\prec_{\kappa} x$
\item $\kappa$ is transitive, i.e., $x\prec_{\kappa}y \land y\prec_{\kappa}z$ $\Rightarrow$ $x\prec_{\kappa}z$.
\end{itemize}
The intuitive interpretation of the relation $x\prec_{\kappa}y$ is that $x$ is ranked before $y$, or $x$ is preferred over $y$.  A pair of candidates are unrelated (incomparable) if $(x\not\prec_{\kappa}y) \land (y\not\prec_{\kappa}x)$. Naturally, the journals are corresponding to the candidates, and a given journal is ranked before all of its descendants in the hierarchy. However, in case of the flow hierarchy only the levels of the hierarchy are given, the ancestor-descendant relations between the journals are not specified. Thus, the flow hierarchy is actually corresponding to a bucket order, where the ``buckets'' are given by the hierarchy levels, and we assume that a journal in a given bucket is preceding all journals in lower buckets, and journals in the same bucket are all equal to each other.

The Kendall-tau distance measure was originally defined for total orders, where all pairs of candidates are comparable. In this case the distance measure is corresponding to the number of inversions needed to convert one total order to the other one. It can be normalised by dividing by the total number of relations, resulting in a value between 0 and 1. In contrast to similarity measures, for an identical pair of total orders, the Kendall-tau distance is 0, while for maximally different total orders it is 1. Here we adapt this concept to the problem of comparing a bucket order (the flow hierarchy) and a partial order (the nested hierarchy). 

The basic idea is to iterate over all possible pairs of journals and compare their ordering in the bucket order and in the partial order. Whenever we observe a mismatch between the two ordering, we increase the distance score $D$ by one. The detailed rules for updating $D$ are the following:
\begin{itemize}
\item $D$ is left unchanged if
\begin{itemize}
\item $x\prec y$ according to the both the bucket order and the partial order, 
\item $x$ and $y$ are unrelated according to the partial order, (are on different branches)
\end{itemize}
\item $D$ is increased by one in all other cases, that is
\begin{itemize}
\item if $x\prec y$ according to the bucket order and $y\prec x$ according to the partial order, 
\item if $x\equiv y $ according to the bucket order, while $x\prec y$ or $y\prec x$ according to the partial order.
\end{itemize} 
\end{itemize}
In order to normalise $D$ we have to divide the obtained result by the maximal number of possible mismatches, which is given by the total number of comparable pairs in the partial order. (I.e., if the given pair is unrelated, $D$ is left unchanged irrespectively to the ordering in the bucket order). 

By applying the above comparison method, our result for the normalised Kendall-tau distance between the flow hierarchy and the nested hierarchy is $D=0.1594$. For comparison we also calculated the mean distance between randomised hierarchies. The randomisation was carried out by simply swapping pairs of journals in a given hierarchy at random, by keeping the structure of the hierarchy (number of levels, number of nodes in a level, etc.) fixed. The average distance and standard deviation was given by $\left< D_{\rm rand}\right>=0.8021\pm 0.0169$. Thus, the examined two hierarchies are significantly closer to each other than expected at random, i.e., the $z$-score for the distance is $-38.03626$.

\section{Comparing the hierarchies with other impact measures}

\begin{figure}[hbt]
\centerline{\includegraphics[width=0.8\textwidth]{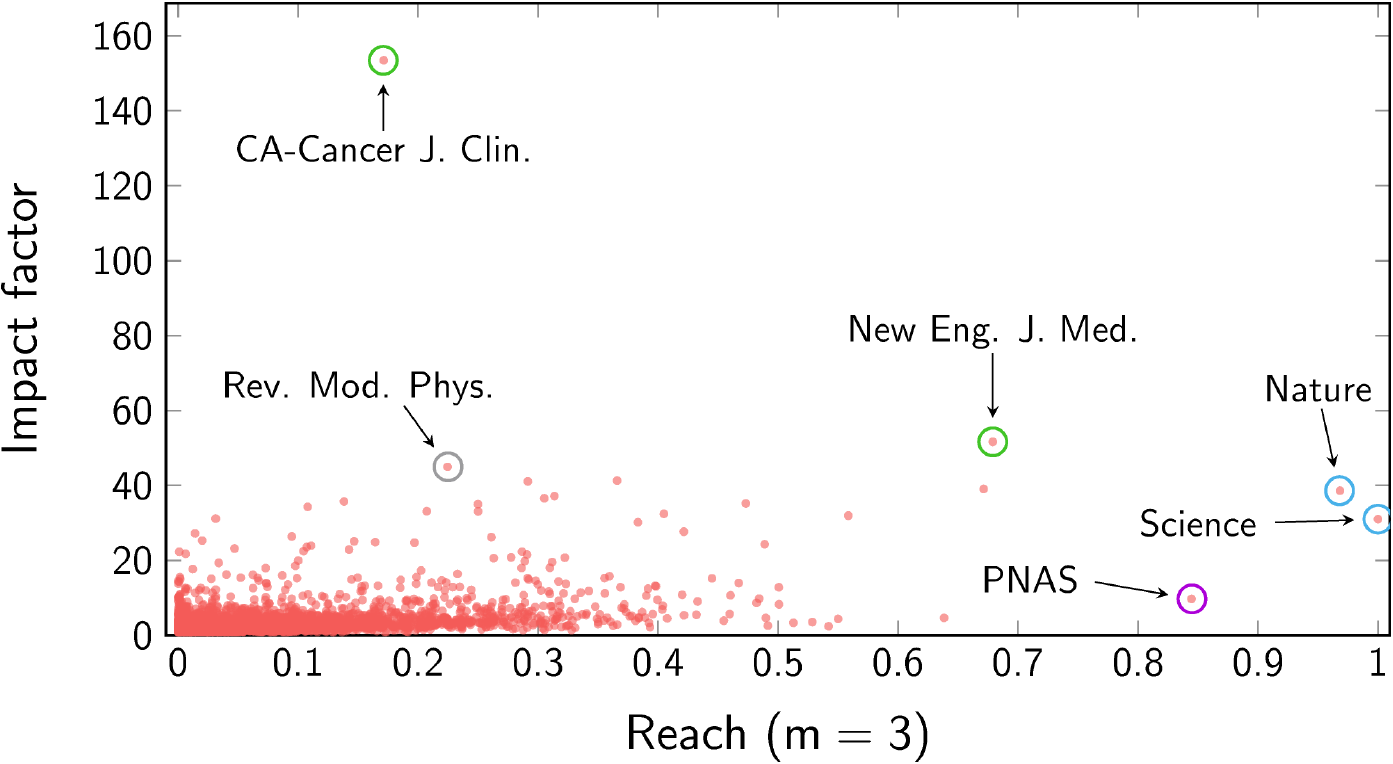}}
\caption{The 2012 journal impact factor as a function of the $m$-reach, $C_m(\mathcal{J})$, given in Eq.(1) in the main paper for the journals in our data base. The $m$-reach was calculated at $m=3$ and was normalised by the largest $m$-reach value in the sample. Journals with the highest impact factor and $m$-reach are highlighted with green and blue circles, respectively. In addition, two more journals appearing high in both the flow- and the nested hierarchy are highlighted in purple and grey.}
\label{fig:reach-vs-if}
\end{figure}

\begin{figure}[h!]
\centerline{\includegraphics[width=0.8\textwidth]{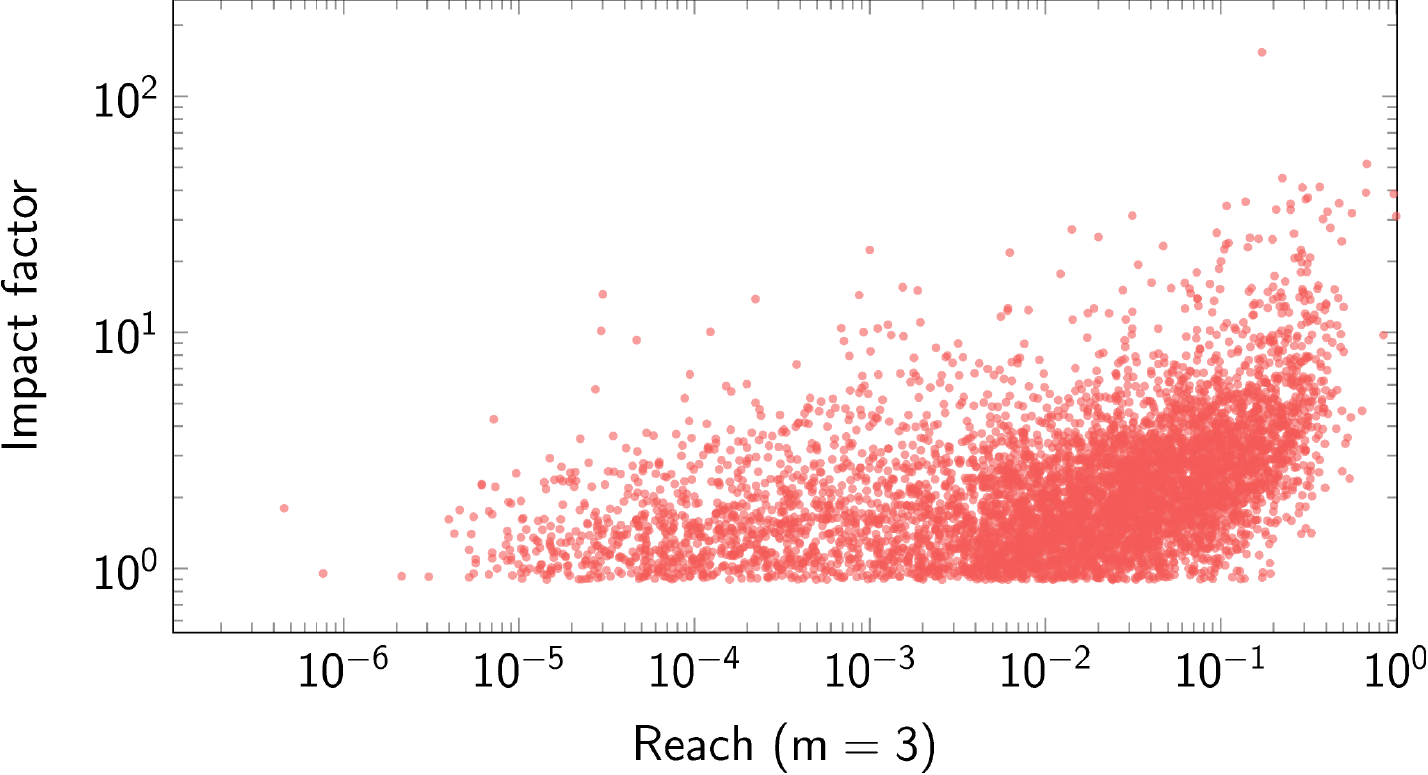}}
\caption{The same scatter plot as in Fig.\ref{fig:reach-vs-if}.\ on logarithmic scale. The 2012 journal impact factor is plotted as a function of the $m$-reach of the journals, normalised by the largest $m$-reach value in the sample.}
\label{fig:reach-vs-if-log}
\end{figure}

We also compare the hierarchies we constructed with more traditional impact measures. The flow hierarchy is defined based on the $m$-reach of the journals given in Eq.(1) in the main paper. This quantity can be directly compared to any traditional impact measure in a simple fashion. Along this line, in Figs.\ref{fig:reach-vs-if}-\ref{fig:reach-vs-if-log}.\ we show the 2012 journal impact factor obtained from Thomson Reuters (2015) as a function of the $m$-reach. The scatter plot suggests moderate correlations, which is supported by the $C_{\rm Pearson}=0.498$ Pearson's correlation coefficient and $C_{\rm Spearman}=0.646$ Spearman's rank correlation coefficient. A few outlier journals are identified in Fig.\ref{fig:reach-vs-if}, e.g., CA: A Cancer Journal for Clinicians has a rather large impact factor, accompanied by a relatively low reach, whereas in contrast PNAS has considerably lower impact factor and a quite large reach. 

\begin{figure}[hbt]
\centerline{\includegraphics[width=0.8\textwidth]{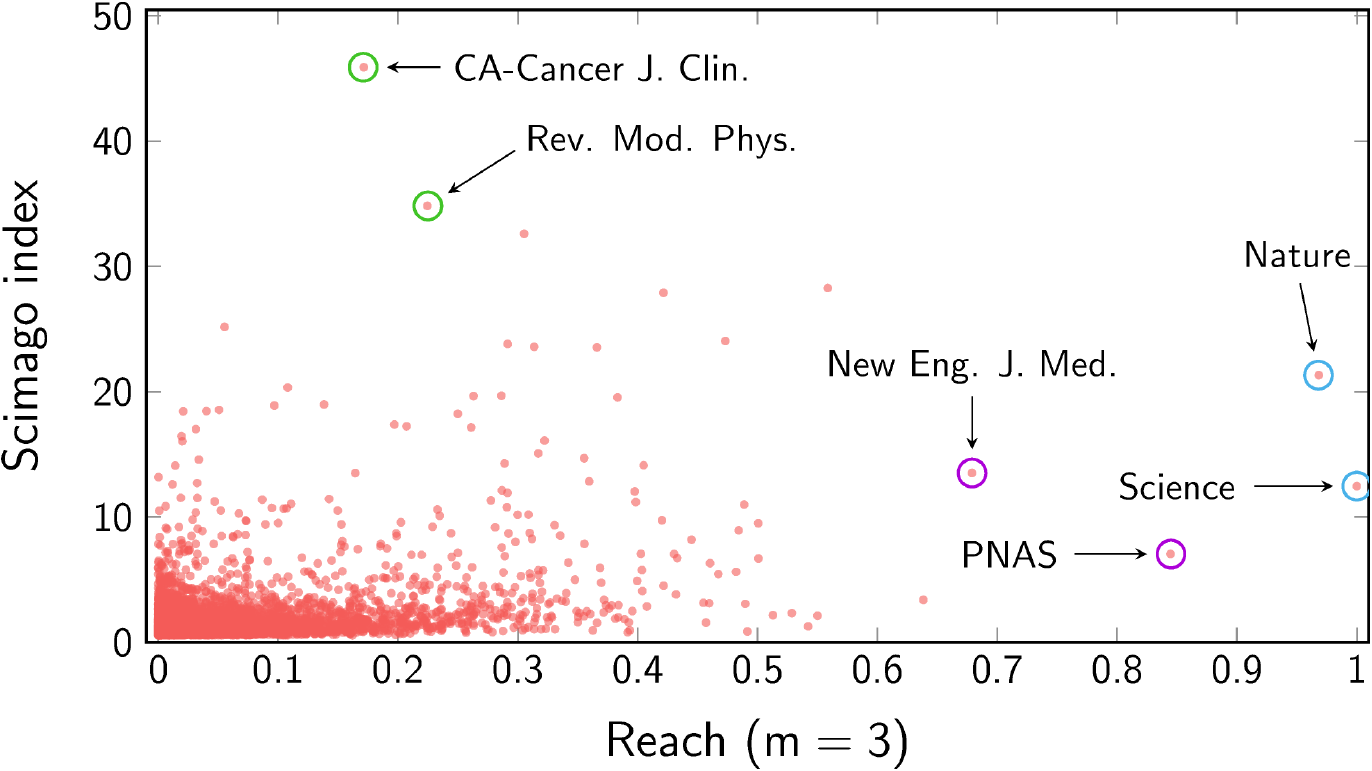}}
\caption{The Scimago Journal Rank as a function of the $m$-reach, $C_m(\mathcal{J})$ for the journals in our data base. The $m$-reach was calculated at $m=3$ and was normalised by the largest $m$-reach value in the sample. Journals with the highest Scimago Journal Rank and $m$-reach are highlighted with green and blue circles, respectively. In addition, two more journals appearing high in both the flow- and the nested hierarchy are highlighted in purple.}
\label{fig:reach-vs-scimago}
\end{figure}

\begin{figure}[h!]
\centerline{\includegraphics[width=0.8\textwidth]{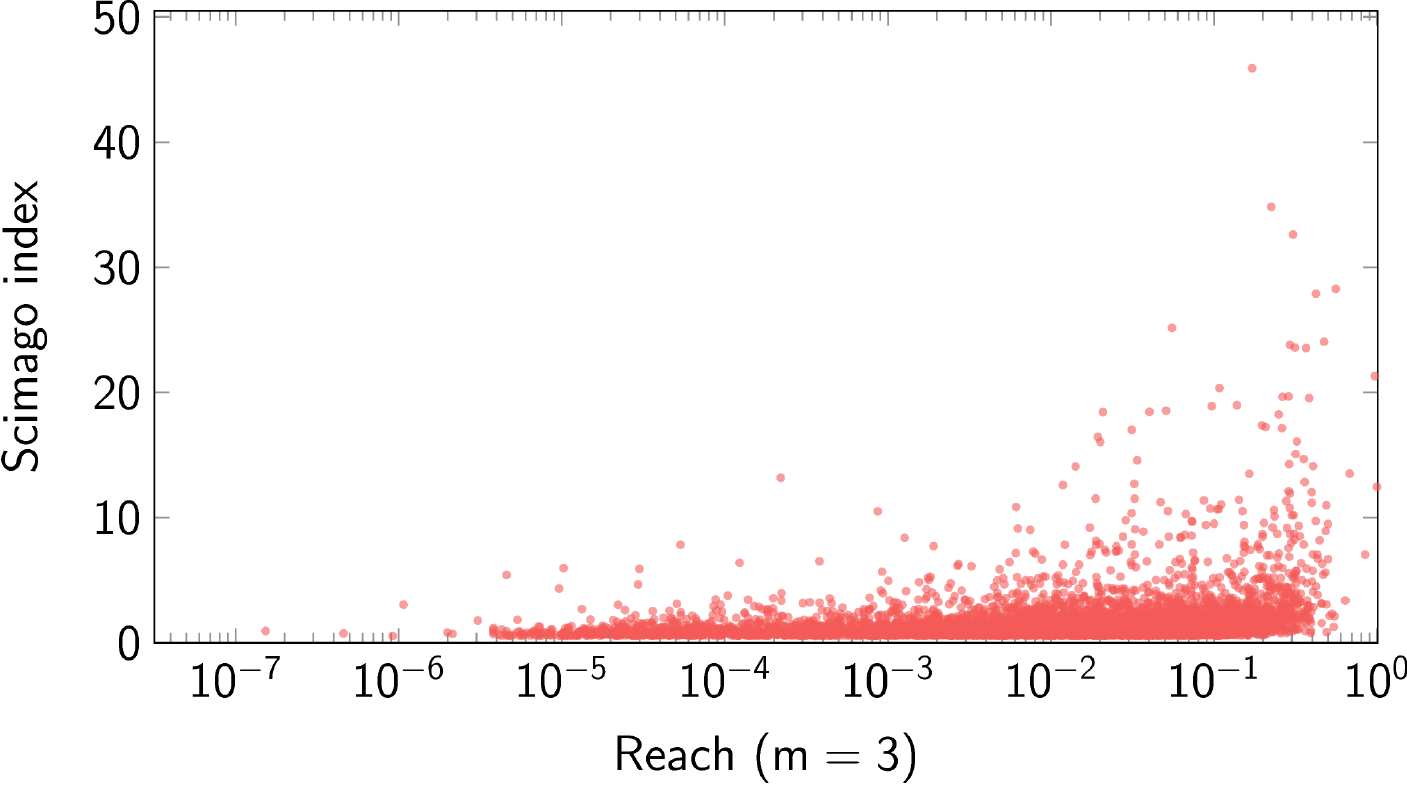}}
\caption{The scatter plot as in Fig.\ref{fig:reach-vs-scimago}.\ on logarithmic scale. The Scimago Journal Rank is plotted as a function of the $m$-reach of the journals  normalised by the largest $m$-reach value in the sample.}
\label{fig:reach-vs-scimago-log}
\end{figure}

In Figs.\ref{fig:reach-vs-scimago}-\ref{fig:reach-vs-scimago-log}.\ we show the latest Scimago Journal Rank \citep{scimago} %(The Scimago Journal \& Country Rank 2015) 
as a function of the $m$-reach. Similarly to the impact factor, the scatter plot is revealing a moderate correlation between the two quantities, in consistency with the $C_{\rm Pearson}=0.460$ Pearson's correlation coefficient and $C_{\rm Spearman}=0.584$ Spearman's rank correlation coefficient obtained for the shown data.

\begin{figure}[hbt]
\centerline{\includegraphics[width=0.8\textwidth]{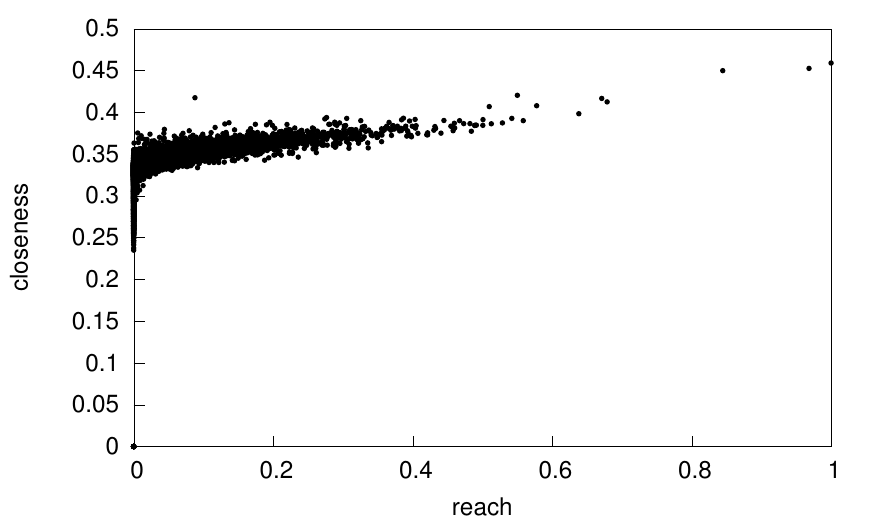}}
\caption{The closeness value of the journals in the citation network as a function of the $m$-reach, $C_m(\mathcal{J})$ for the journals in our data base. The $m$-reach was calculated at $m=3$ and was normalised by the largest $m$-reach value in the sample.}
\label{fig:reach-vs-closeness}
\end{figure}

Finally, in Fig.\ref{fig:reach-vs-closeness}.\ we show the closeness of the journals calculated in the aggregated citation network as a function of the $m$-reach. Similarly to the previous impact measures, we can observe a moderate correlation between the two quantities. The corresponding Pearson's correlation coefficient is given by $C_{\rm Pearson}=0.466$, and the Spearman's rank correlation coefficient is equal to $C_{\rm Spearman}=0.823$.
\begin{figure}[h!]
\centerline{\includegraphics[width=0.8\textwidth]{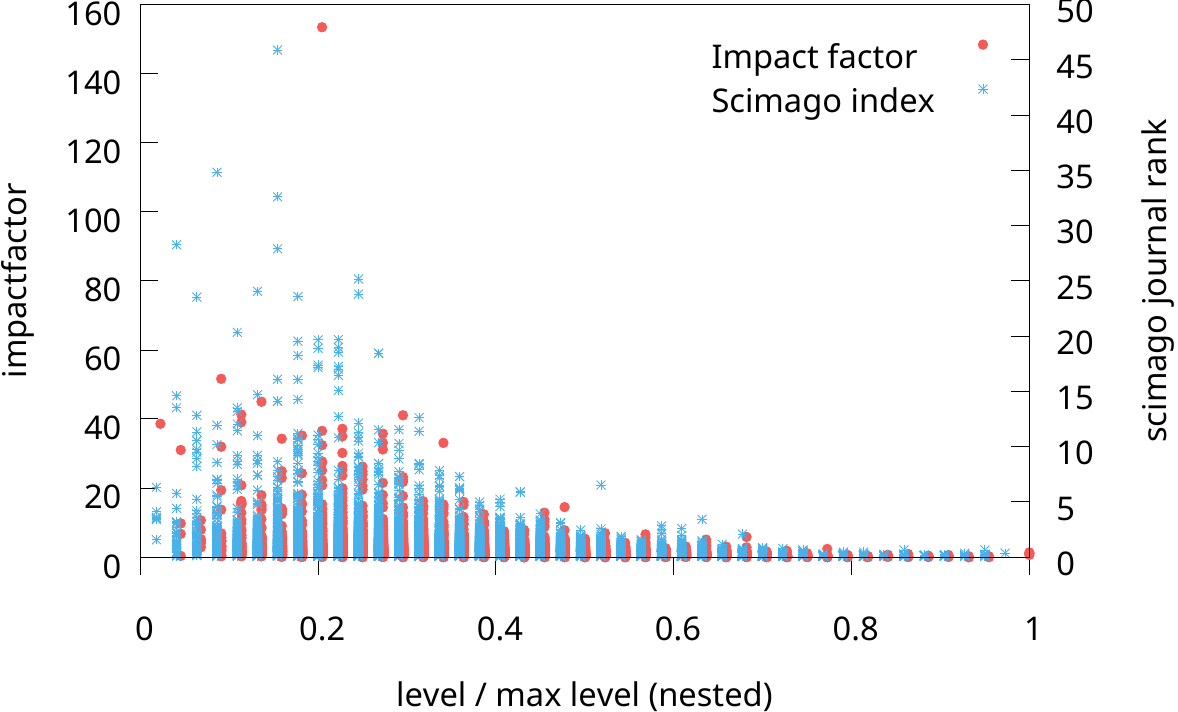}}
\caption{ The 2012 journal impact factor (red circles) and the Scimago Journal Ranks (blue crosses) as a function of the level depth in the nested hierarchy, where the level depth was normalised by the maximal level depth. Note that the scale for the Scimago Journal Ranks is given on the vertical axis on the right.}
\label{fig:nested-vs-if}
\end{figure}

In addition, we may also apply the concept of the Kendall-tau distance defined in Sect.\ref{sect:Kendall}.\ for comparing the impact factor or the Scimago Journal Rank with the flow hierarchy. The resulting $D=0.072$ for the distance between the ranking by the impact factor and the flow hierarchy, and also the $D=0.086$ between the Scimago Journal Rank and the flow hierarchy are corresponding to larger distances than the $D$ obtained for the distance between the nested hierarchy and the flow hierarchy. However they are also significantly lower compared to the distance between randomised rankings, providing $\left< D_{\rm rand}\right>=0.198\pm 0.003$ both for the impact factor the Scimago Journal Rank. We note that there are a several ties in both the journal impact factors and the Scimago Journal Ranks, therefore, the structure of the corresponding partial orders is not exactly like a linear chain. Accordingly, the distance between the randomised flow hierarchy and the randomised impact factor rankings can be different from the distance between the randomised flow hierarchy and the randomised Scimago Journal Ranks in theory.
\begin{figure}[h!]
\centerline{\includegraphics[width=0.8\textwidth]{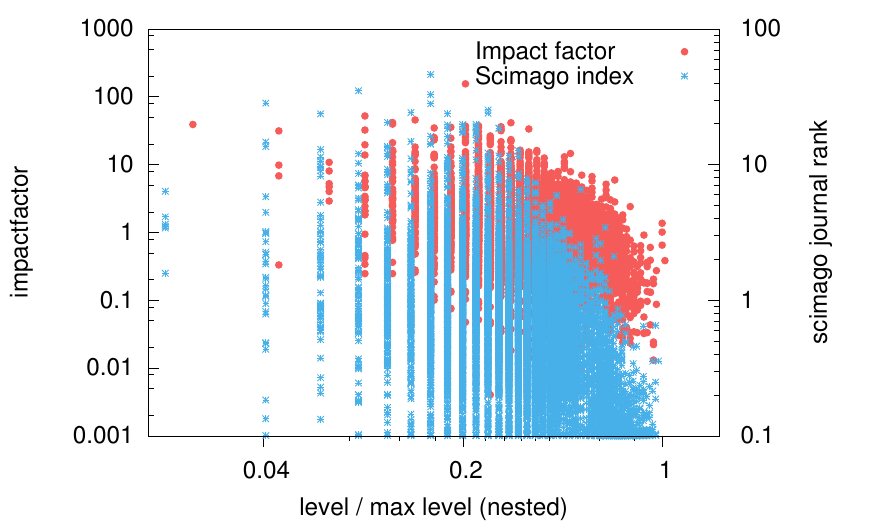}}
\caption{ The same scatter plot as in Fig.\ref{fig:nested-vs-if}. on logarithmic scale. The 2012 journal impact factor is shown by red circles, while the Scimago Journal Ranks are marked by blue crosses. Both impact measures are plotted as a function of the normalised level depth in the nested hierarchy. Note that the scale for the Scimago Journal Ranks is given on the vertical axis on the right.}
\label{fig:nested-vs-if-log}
\end{figure}

\begin{figure}[hbt]
\centerline{\includegraphics[width=0.8\textwidth]{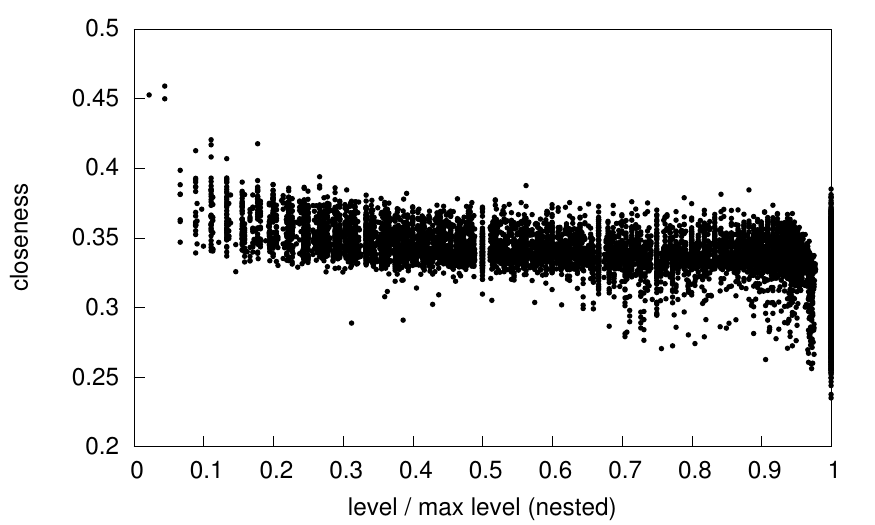}}
\caption{The closeness value of the journals in the citation network as a function of the level depth in the nested hierarchy, where the level depth was normalised by the maximal level depth.}
\label{fig:nested-vs-closeness}
\end{figure}

In case of the nested hierarchy in Figs.\ref{fig:nested-vs-if}-\ref{fig:nested-vs-if-log}.\ we show the journal impact factors and the Scimago Journal Ranks as functions of the level depth. These scatter plots indicate a moderate negative correlation between the traditional impact measures and the level depth as expected: Journals with high impact factor and high Scimago Journal Rank tend to be placed on the top levels, whereas the impact measures at the lower levels in the hierarchy seem to be lower on average. This is supported by the $C_{\rm Pearson}=-0.272$ Pearson correlation and the $C_{\rm Spearman}=-0.211$ Spearman's rank correlation coefficient between the impact factor and the nested hierarchy level depth, and the $C_{\rm Pearson}=-0.289$ and $C_{\rm Spearman}=-0.314$ correlation coefficients between the Scimago Journal Rank and the nested hierarchy level depth. In addition, in Fig.\ref{fig:nested-vs-closeness}.\ we show the closeness value of the journals in the aggregated citation network as a function of their normalised level depth in the nested hierarchy. According to the plot, the two quantities show moderate correlation, with a corresponding Pearson's correlation coefficient of $C_{\rm Pearson}=-0.449$.

Although the magnitude of these correlation coefficients is somewhat smaller than in case of the flow hierarchy, the nested hierarchy is still about 10 times closer to the ranking by the impact factor or by the Scimago Journal Ranks according to the Kendall-tau distance: We obtained a $D=0.065$ distance between the nested hierarchy and the ranking according to the impact factor, in contrast to the $\left< D_{\rm rand}\right>=0.60\pm 0.025$ average and standard deviation for the distance between two corresponding randomised rankings. In case of the Scimago Journal Rank, the Kendall-tau distance is $D=0.057$, while the average and standard deviation for the randomised rankings is $\left< D_{\rm rand}\right>=0.460 \pm0.015$.

\end{document}